\documentclass[aps,prb,twocolumn,showpacs,amsmath,amssymb,superscriptaddress]{revtex4}
\usepackage{graphicx}
\usepackage{amssymb}
\usepackage{natbib}
\usepackage{color}
\usepackage{latexsym}
\usepackage{amsmath}
\usepackage{endnotes}
\usepackage{lmodern}
\input{epsf}

\newcommand{\MM}{\mathbf{M}}

\newcommand{\beq}{\begin{eqnarray}}
\newcommand{\eeq}{\end{eqnarray}}
\newcommand{\ud}{\mathrm{d}}
\newcommand{\vS}{\mathbf{S}}

\newcommand{\eps}{\varepsilon}
\newcommand{\SrFeNi}{SrFe$_{1.97}$Ni$_{0.03}$As$_2^{\hphantom{x}}$}
\newcommand{\BaFeNi}{BaFe$_{1.97}$Ni$_{0.03}$As$_2^{\hphantom{x}}$}
\newcommand{\BaFeAs}{BaFe$_2$As$_2^{\hphantom{X}}$}

\begin{document}

\title{Impact of Uniaxial Pressure on Structural and Magnetic Phase Transitions in Electron-Doped
Iron Pnictides}
\author{Xingye Lu}
\affiliation{ Department of Physics and Astronomy \& Rice Center for Quantum Materials,
Rice University, Houston, Texas 77005, USA }
\affiliation{Beijing National Laboratory for Condensed Matter
Physics, Institute of Physics, Chinese Academy of Sciences, Beijing
100190, China}

\author{Kuo-Feng Tseng}
\affiliation{Max-Planck-Institut f$\ddot{u}$r Festk$\ddot{o}$rperforschung, Heisenbergstrasse 1, D-70569 Stuttgart, Germany}
\affiliation{Max Planck Society Outstation at the Forschungsneutronenquelle Heinz Maier-Leibnitz (MLZ), D-85747 Garching, Germany}

\author{T. Keller}
\affiliation{Max-Planck-Institut f$\ddot{u}$r Festk$\ddot{o}$rperforschung, Heisenbergstrasse 1, D-70569 Stuttgart, Germany}
\affiliation{Max Planck Society Outstation at the Forschungsneutronenquelle Heinz Maier-Leibnitz (MLZ), D-85747 Garching, Germany}

\author{Wenliang Zhang}
\affiliation{Beijing National Laboratory for Condensed Matter
Physics, Institute of Physics, Chinese Academy of Sciences, Beijing
100190, China}

\author{Ding Hu}
\affiliation{Beijing National Laboratory for Condensed Matter
Physics, Institute of Physics, Chinese Academy of Sciences, Beijing
100190, China}

\author{Yu Song}
\affiliation{ Department of Physics and Astronomy \& Rice Center for Quantum Materials,
Rice University, Houston, Texas 77005, USA }

\author{Haoran Man}
\affiliation{ Department of Physics and Astronomy \& Rice Center for Quantum Materials,
Rice University, Houston, Texas 77005, USA }

\author{J. T. Park}
\affiliation{Heinz Maier-Leibnitz Zentrum (MLZ), Technische Universit$\ddot{a}$t M$\ddot{u}$nchen, 85748 Garching, Germany}

\author{Huiqian Luo}
\affiliation{Beijing National Laboratory for Condensed Matter
Physics, Institute of Physics, Chinese Academy of Sciences, Beijing
100190, China}

\author{Shiliang Li}
\affiliation{Beijing National Laboratory for Condensed Matter
Physics, Institute of Physics, Chinese Academy of Sciences, Beijing
100190, China}
\affiliation{Collaborative Innovation Center of Quantum Matter, Beijing, China}

\author{Andriy H. Nevidomskyy}
\affiliation{ Department of Physics and Astronomy \& Rice Center for Quantum Materials,
Rice University, Houston, Texas 77005, USA }

\author{Pengcheng Dai}
\email{pdai@rice.edu}
\affiliation{ Department of Physics and Astronomy \& Rice Center for Quantum Materials,
Rice University, Houston, Texas 77005, USA }

\date{\today}

\begin{abstract}
We use neutron resonance spin echo and Larmor diffraction to study the effect of uniaxial pressure on the tetragonal-to-orthorhombic
structural ($T_s$) and antiferromagnetic (AF) phase transitions in iron pnictides
BaFe$_{2-x}$Ni$_{x}$As$_{2}$ ($x=0,0.03,0.12$), SrFe$_{1.97}$Ni$_{0.03}$As$_2$, and BaFe$_2$(As$_{0.7}$P$_{0.3}$)$_2$.
In antiferromagnetically ordered BaFe$_{2-x}$Ni$_{x}$As$_{2}$ and SrFe$_{1.97}$Ni$_{0.03}$As$_2$ with
$T_N$ and $T_s$ ($T_N\leq T_s$),
a uniaxial pressure necessary to detwin the sample also increases $T_N$,
 smears out the structural transition, and induces an orthorhombic lattice distortion at all temperatures.
By comparing temperature and doping dependence of the pressure induced lattice parameter changes with
the elastoresistance and nematic susceptibility
obtained from transport and ultrasonic measurements,
we conclude that the in-plane resistivity anisotropy found in the paramagnetic state of electron underdoped iron pnictides
depends sensitively on the nature of the magnetic phase transition and
a strong coupling between the uniaxial pressure induced lattice distortion and
electronic nematic susceptibility.
\end{abstract}

\maketitle

\section{INTRODUCTION}

The parent compounds of iron pnictide superconductors such as
BaFe$_2$As$_2$ and SrFe$_2$As$_2$ exhibit a tetragonal-to-orthorhombic structural transition
at $T_s$ followed by development of collinear antiferromagnetic (AF) order along the $a$-axis of the
orthorhombic lattice below $T_N$ [left inset in Fig. 1(a) and $T_s\approx T_N$] \cite{kamihara,cruz,qhunag,JZhao,mgkim,pcdai}.
Upon electron-doping via partially substituting Fe by Co or Ni to form BaFe$_{2-x}T_{x}$As$_{2}$ ($T=$ Co, Ni),
the nearly coupled structural and magnetic phase transitions in BaFe$_2$As$_2$ become two separate
second order phase transitions at $T_s$ and $T_N$ ($T_s>T_N$) that decrease in temperature
with increasing $x$ \cite{CLester2009,SNandi,hqluo,xylu13}. On the other hand, the coupled first order
structural and magnetic phase transitions in SrFe$_2$As$_2$ \cite{JZhao}, while decreasing in temperature
with increasing $x$ in SrFe$_{2-x}T_{x}$As$_{2}$, remain coupled first
order transitions leading up to superconductivity \cite{RWHu}.

Because the structural and magnetic phase transitions
in BaFe$_{2-x}T_{x}$As$_{2}$ and SrFe$_{2-x}T_{x}$As$_{2}$
occur below room temperature,
iron pnictides in the orthorhombic AF ground state will form twin domains with
AF Bragg peaks appearing at the in-plane $(\pm 1,0)$ and $(0,\pm 1)$
positions in reciprocal space [right inset in Fig. 1(a)] \cite{pcdai}. To probe the intrinsic electronic properties
of these materials, one can apply uniaxial pressure along one axis of the
orthorhombic lattice to obtain single domain samples \cite{chu10,matanatar,fisher,ishida}.  Indeed, transport measurements on uniaxial
pressure detwinned electron-doped BaFe$_{2-x}T_{x}$As$_{2}$ ($T=$ Co, Ni) reveal in-plane resistivity anisotropy
in the AF state that persists to temperatures above the
zero-pressure $T_N$ and $T_s$ \cite{chu10,matanatar,fisher,ishida}.  On the other hand, similar transport measurements on uniaxial pressured
detwinned SrFe$_{2-x}T_{x}$As$_{2}$ ($T=$ Co, Ni) indicate vanishingly small resistivity anisotropy
at temperatures above the zero pressure coupled $T_N$ and $T_s$ \cite{Blomberg,Blomberg12,SDDas}.  Figure 1(b) compares temperature dependence
of the resistivity anisotropy [defined as $\Delta\rho = (\rho_b-\rho_a)/(\rho_b+\rho_a)$, where
$\rho_a$ and $\rho_b$ are resistivity along the $a$ and $b$ axis of the orthorhombic lattice, respectively] obtained under 20 MPa uniaxial pressure for BaFe$_2$As$_2$, BaFe$_{1.97}$Ni$_{0.03}$As$_2$, SrFe$_{1.97}$Ni$_{0.03}$As$_2$, and SrFe$_2$As$_2$.  Consistent with earlier works \cite{chu10,matanatar,fisher,ishida,Blomberg,Blomberg12,SDDas}, we find that resistivity anisotropy is much larger
in BaFe$_2$As$_2$ and BaFe$_{1.97}$Ni$_{0.03}$As$_2$ at temperatures above $T_N$.

Although resistivity anisotropy in the paramagnetic state of the iron pnictides under applied uniaxial pressure
suggests the presence of an electronic nematic phase that breaks
the in-plane fourfold rotational symmetry ($C_4$) of the underlying
tetragonal lattice \cite{fradkin,chu12,Kasahara,Gallais,HHKuo2014,HHKuo2015}, much is unclear about the
microscopic origin of the in-plane resistivity anisotropy and electronic nematic phase
 \cite{RMfernandes11,fernandes14,CCL,jphu,si,myi,cclee,kruger,lv,ccchen,valenzeula,wang15,Kantani,anna}.
Since neutron scattering experiments reveal that
uniaxial pressure necessary to detwin the sample also increases $T_N$ of the system,
the observed in-plane resistivity anisotropy above the zero pressure $T_N$ and $T_s$
may arise from the increased $T_N$ and intrinsic anisotropic nature of
the collinear AF phase \cite{Cdhital,Cdhital14}. Furthermore, while it is generally assumed that
the uniaxial pressure for sample detwinning has negligible effect on the lattice parameters
of the iron pnictides \cite{chu10,matanatar,fisher,ishida}, the precise effect of uniaxial pressure on structural distortion of
these materials is unknown.  From neutron extinction effect measurements,
a uniaxial pressure is suggested to push structural fluctuations
related to the orthorhombic distortion to a
temperature well above the zero-pressure value of $T_s$ \cite{xyscience}, similar to  the effect on the resistivity anisotropy \cite{chu10,matanatar,fisher,ishida}.
To understand the microscopic origin of the in-plane resistivity anisotropy
in the paramagnetic state \cite{chu10,matanatar,fisher,ishida},
it is important to establish the effect of
a uniaxial pressure on the magnetic and structural phase transitions of
BaFe$_{2-x}T_{x}$As$_{2}$ and SrFe$_{2-x}T_{x}$As$_{2}$, and determine if the electronic anisotropy in
the paramagnetic tetragonal phase of iron pnictides is intrinsic \cite{mpallan,Rosenthal}, or
entirely due to the symmetry breaking uniaxial pressure applied to the materials \cite{Mirri,HRMan}. It is also important to deduce what role the nature of the AF transition plays in the nematic susceptibility \cite{Blomberg12,RMfernandes11,fernandes14,anna} and how the latter depends on the uniaxial pressure.

\begin{figure}[t]
\includegraphics[width=7cm]{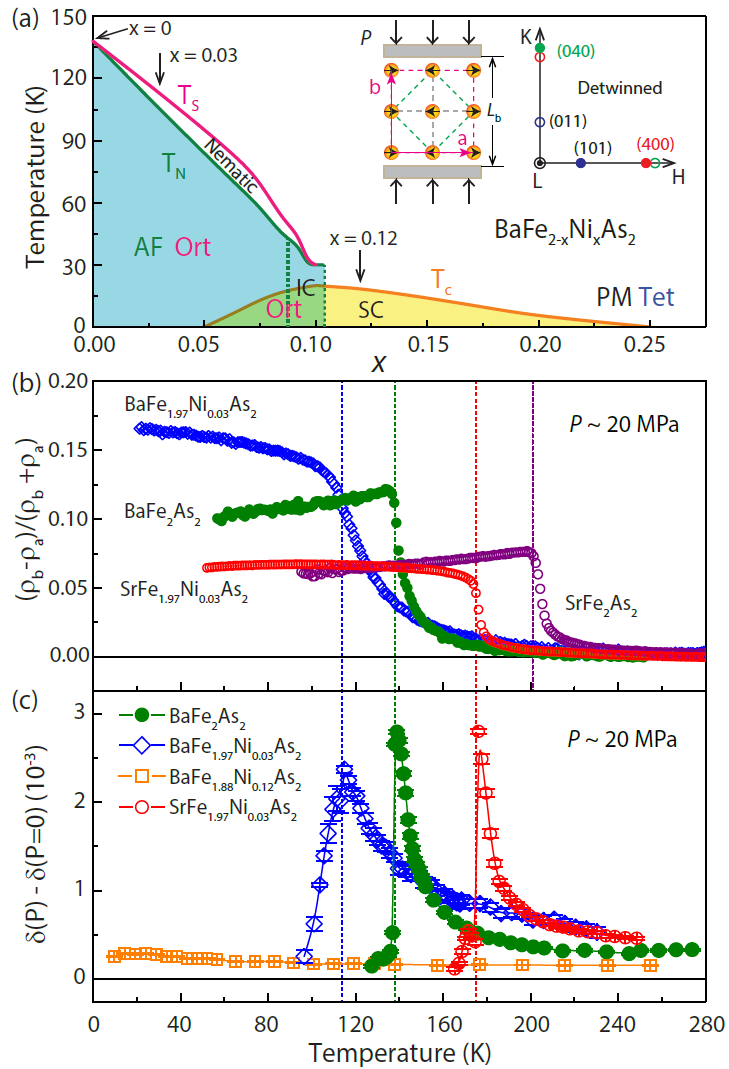}
\caption{(Color online) (a) The schematic electronic phase diagram of BaFe$_{2-x}$Ni$_x$As$_2$
with arrows marking $x=0, 0.03$ and $0.12$ samples described in
the present study. The AF, PM, Ort, Tet, IC, SC are antiferromagnetic,
paramagnetic, orthorhombic, tetragonal, incommensurate,
and superconducting states, respectively \cite{xylu13}.
The left inset shows the direction of the applied uniaxial pressure (marked by the vertical arrows)
and the spin arrangements of Fe in the AF ordered iron pnictides, where
$a$ and $b$ are the orthorhombic axes. The right inset shows the corresponding
reciprocal lattice. All the marked positions have AF or nuclear Bragg peaks for a twinned sample,
while the positions marked by open symbols have vanishing scattering intensity
for a detwinned sample. (b) Temperature dependence of the resistivity anisotropy
for BaFe$_{2-x}$Ni$_{x}$As$_{2}$ and SrFe$_{2-x}$Ni$_{x}$As$_{2}$ ($x=0,0.03$) under $P\approx 20$ MPa.
(c) Summary of temperature dependence of the
uniaxial pressure induced lattice distortion at $P=20$ MPa
[$\delta(P=20\ {\rm MPa})-\delta(P=0\ {\rm MPa})$] for
BaFe$_{2-x}$Ni$_{x}$As$_{2}$ ($x=0,0.03,0.12$)
and SrFe$_{1.97}$Ni$_{0.03}$As$_2$.
The actual data for $x=0.03, 0.12$ are normalized to 20 PMa assuming a linear relationship between
uniaxial pressure and $\delta$.  Uniaxial pressure induced lattice distortion vanishes rapidly below $T_N$
marked by the vertical dashed lines in (b) and (c).
}
\end{figure}

In this paper, we use neutron resonance spin echo (NRSE) \cite{Mezei,TKeller} and Larmor diffraction \cite{Bayrakci} to study the
effect of uniaxial pressure on the
structural and magnetic phase transitions in electron doped iron pnictides
BaFe$_{2-x}$Ni$_{x}$As$_{2}$ with $x=0,0.03, 0.12$ \cite{hqluo,xylu13} and SrFe$_{1.97}$Ni$_{0.03}$As$_2$ \cite{SDDas},
and in the isovalently doped BaFe$_2$(As$_{0.7}$P$_{0.3}$)$_2$ \cite{Dhu}.
While the
underdoped BaFe$_{1.97}$Ni$_{0.03}$As$_{2}$ ($T_N=109$ K and $T_s=114$ K)
exhibits a second-order
AF transition below $T_s$, SrFe$_{1.97}$Ni$_{0.03}$As$_2$
has coupled first-order structural and magnetic phase transitions at $T_N=T_s\approx 175$ K \cite{supplementary}.
The electron overdoped BaFe$_{1.88}$Ni$_{0.12}$As$_{2}$ ($T_c=18.6$ K) and isovalently doped
BaFe$_2$(As$_{0.7}$P$_{0.3}$)$_2$ ($T_c=30$ K) have a
paramagnetic tetragonal structure at all temperatures without static AF order.
Figure 1(c) summarizes the key experimental result of the present work, where
the temperature dependences of
the uniaxial pressure induced orthorhombic lattice distortion $\delta(P=20\ {\rm MPa})-\delta(P=0\ {\rm MPa})$ are determined using neutron Larmor diffraction for BaFe$_2$As$_2$, BaFe$_{1.97}$Ni$_{0.03}$As$_{2}$, SrFe$_{1.97}$Ni$_{0.03}$As$_2$, and
BaFe$_{1.88}$Ni$_{0.12}$As$_{2}$ [we defined the lattice distortion
$\delta=(a-b)/(a+b)$ with $a$ and $b$ being the orthorhombic
lattice parameters]. Remarkably, the magnitude of our determined structural nematic susceptibility $\ud \delta/\ud P \propto \delta(P) - \delta(0)$ in Figure 1(c) is comparable in all three materials that have a structural phase transition, unlike the very different values of the resistivity anisotropy displayed in Figure 1(b).
Comparing these results with
those of the elastoresistance and nematic susceptibility obtained from transport  \cite{chu12,HHKuo2014,HHKuo2015}
and from elastic shear modulus/ultrasound spectroscopy measurements \cite{Fernandes10,Yoshizawa,anna}, we conclude that
the resistivity anisotropy in the paramagnetic phase of the iron pnictides
depends sensitively on whether the underlying magnetic phase transition is first or second order.
We also find a strong coupling between the uniaxial pressure induced lattice distortion and the electronic nematic susceptibility, and have to be cautious in directly relating resistivity anistropy to the nematic order parameter
in the iron pnictides.

\begin{figure}[t]
\includegraphics[width=7cm]{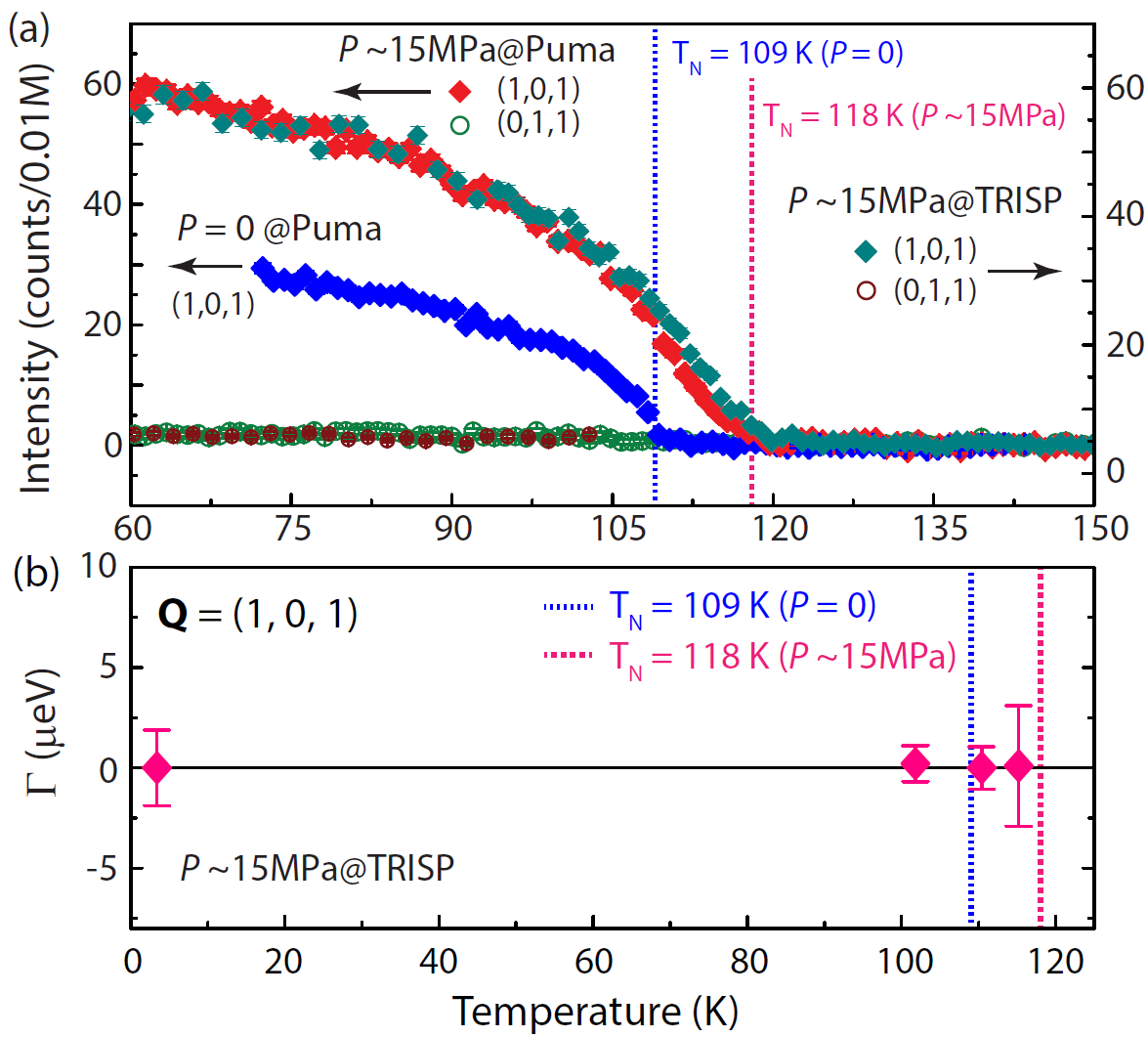}
\caption{ (Color online)
(a) Magnetic order parameters
at $\textbf{Q}=(1,0,1)$ for the zero ($P=0$)
and uniaxial pressured ($P\sim15$MPa) BaFe$_{1.97}$Ni$_{0.03}$As$_2$.
$T_N$ is $109$K for an unpressured sample (blue diamonds).
Upon applying uniaxial pressure of $P\sim15$ MPa,  the $T_N$ is enhanced to $118$K
and the sample becomes $100\%$ detwinned as seen by PUMA  and TRISP measurements.
(b) The energy line width (Half-Width-at-Half-Maximum, $\Gamma$) of the magnetic Bragg peak $\textbf{Q}=(1,0,1)$
measured by NRSE using TRISP for BaFe$_{1.97}$Ni$_{0.03}$As$_2$. The blue and red dashed lines indicate $T_N$ in $P=0$ and $15$ MPa
unaixial pressure, respectively. The slight larger errors of $\Gamma$ near $T_N$ is due to
low statistics data.
}
 \end{figure}

\section{RESULTS}

\subsection{Experimental Results}
Our experiments were carried out using conventional thermal triple-axis spectrometer PUMA and
three axes spin echo spectrometer (TRISP) at
the Forschungsneutronenquelle Heinz Maier-Leibnitz (MLZ), Garching, Germany.
The principles of NRSE and Larmor diffraction are described elsewhere \cite{supplementary}.
Single crystals of BaFe$_{2-x}$Ni$_{x}$As$_{2}$, SrFe$_{2-x}$Ni$_{x}$As$_2$,
and BaFe$_2$(As$_{0.7}$P$_{0.3}$)$_2$  were grown by self-flux method as described before \cite{Dhu,ycchen}.
We define the momentum transfer ${\bf Q}$ in the three-dimensional reciprocal space in \AA$^{-1}$
as $\textbf{Q}=H\textbf{a}^\ast+K\textbf{b}^\ast+L\textbf{c}^\ast$, where $H$, $K$, and $L$ are Miller indices and
${\bf a}^\ast=\hat{{\bf a}}2\pi/a$, ${\bf b}^\ast=\hat{{\bf b}}2\pi/b$,
${\bf c}^\ast=\hat{{\bf c}}2\pi/c$ with  $a\approx b\approx 5.6$ \AA, and $c=12.96$ \AA\ for BaFe$_{2-x}$Ni$_{x}$As$_{2}$.
In this notation, the AF Bragg peaks should occur at $(\pm 1,0,L)$ ($L=1,3,5,\cdots$)
positions in reciprocal space of a completely detwinned sample [right inset in Fig. 1(a)].
For neutron scattering experiments, single crystals are aligned
in either the $[H,K,H+K]$ \cite{xyscience} or $[H,K,0]$ zone.

We first discuss the effect of uniaxial pressure on the collinear AF order in BaFe$_{2-x}T_{x}$As$_{2}$.  In previous neutron
scattering work on BaFe$_{2-x}$Co$_{x}$As$_{2}$, the N$\rm \acute{e}$el temperature ($T_N$) was found to be pushed to higher temperature under uniaxial strain field, forming a broader magnetic transition \cite{Cdhital,Cdhital14}. Moreover, it seems that the increase in $T_N$ depends on the annealing condition \cite{xyscience,YSong13}. Although the $T_N$ enhancement was attributed to uniaxial strain aligned fluctuating magnetic domains, the effect of uniaxial pressure on the ordered moment remains elusive and the nature of the $T_N$ enhancement is still under debate \cite{Cdhital14,xyscience,YSong13}.

By aligning single crystals in the $[1,0,1]\times [0,1,1]$ scattering plane \cite{xyscience}, we were able to determine $T_N$, detwinning ratio, as well as the ordered
moment of the system under zero and finite uniaxial pressures. Figure 2(a) shows temperature dependence of the
$(1,0,1)$ and $(0,1,1)$ magnetic scattering intensity for BaFe$_{1.97}$Ni$_{0.03}$As$_{2}$ obtained
using PUMA [left axis in Fig. 2(a)] and TRISP (right axis). The two sets of data are in excellent quantitative agreement with each other.
Under the applied uniaxial pressure of $P\approx 15$~MPa, the N\'eel temperature increases from $T_N\approx 109$ K (at $P=0$) to $T_N\approx 118$ K. The magnetic scattering intensity [Fig. 2(a)] in the $(1,0,1)$ peak becomes approximately twice as large as in the twinned sample, whereas the
$(0,1,1)$ peak vanishes,  suggesting that the sample is completely detwinned and the
applied uniaxial pressure does not significantly affect the ordered moment.

To test whether the $T_N$ increase is an intrinsic feature of the system, we note that
the magnetic order parameter under uniaxial pressure has a round tail around $T_N$ \cite{Cdhital,Cdhital14}, suggesting
that the $T_N$ enhancement could arise from enhanced slow spin dynamics (critical scattering) under inhomogeneous uniaxial strain field and cannot be resolved by conventional triple-axis neutron diffraction due to its coarse energy resolution ($\Delta E\approx 0.3-1$ meV). To clarify the nature of the increase in $T_N$, we have measured the energy line-width ($\Gamma \ge 0$, [see Fig. 2(b)]) of the quasielastic scattering for magnetic reflection $(1, 0, 1)$ using high energy resolution ($\Delta E\approx 1$ $\mu$eV) NRSE at TRISP \cite{xylu_trisp}. As seen in Figure 2(b), the $\Gamma$ at all measured temperatures are resolution limited, indicating that the increase in magnetic scattering intensity below $T_N\approx 118$ K is elastic ($\Gamma\le 1\mu$eV), and an intrinsic nature of the system.

\begin{figure}[htbp]
\includegraphics[width=7cm]{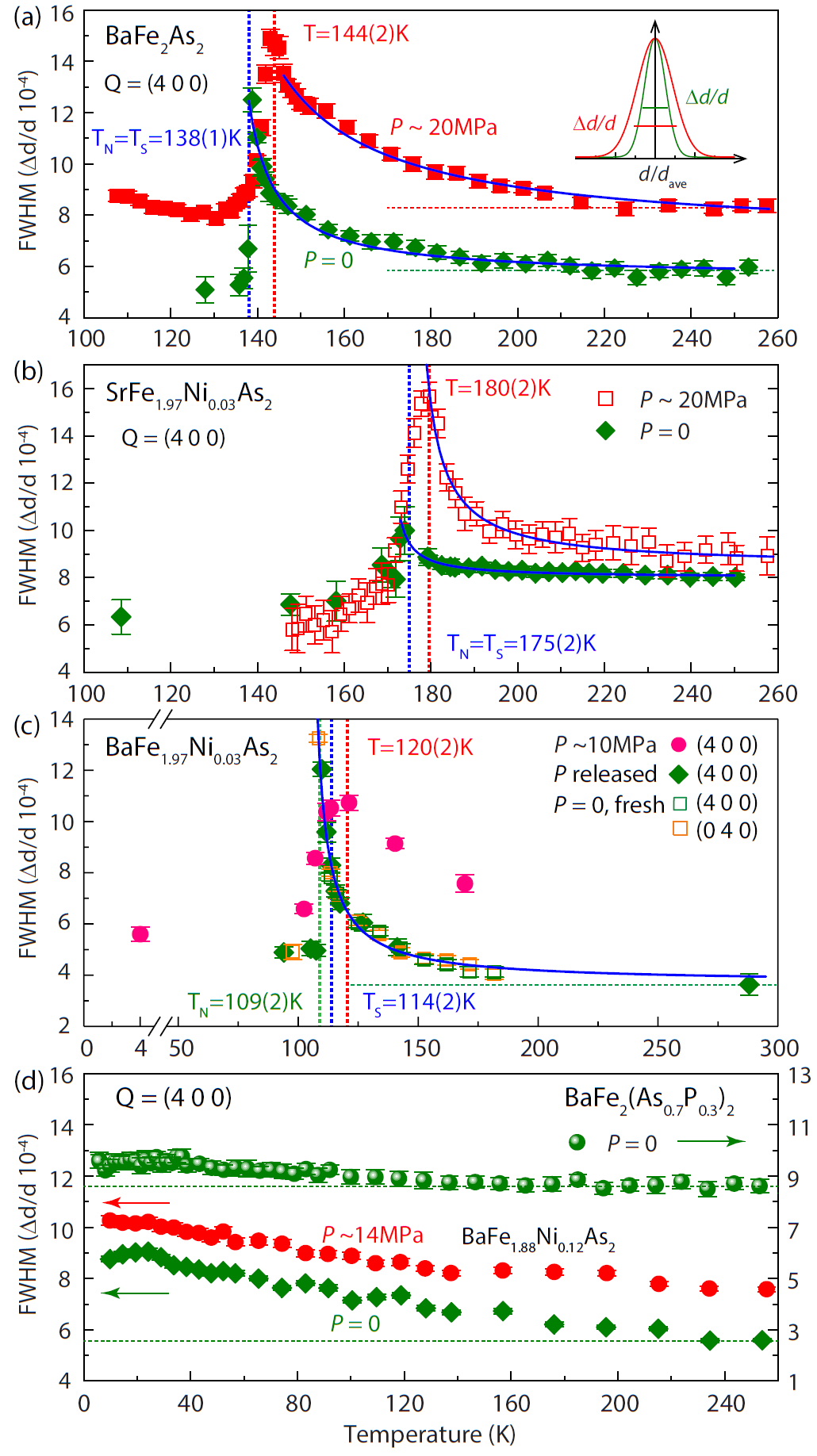}
\caption{ (Color online)
Temperature dependence of FWHM of
$\Delta d/d$ in sev eral iron pnictides under different uniaxial pressures
obtained from neutron Larmor diffraction experiments \cite{supplementary}.
(a) Temperature dependence of
$\Delta d/d$ in FWHM for the $(4,0,0)$ Bragg reflection of BaFe$_2$As$_2$ at $P=0$ and 20 MPa.  The solid line
above $T_N$ is a fit using Curie-Wiess formalism [${\rm FWHM}(T) = A/(T-T_1)+B$, where $A, B,$ and $T_1$ are fitting parameters].
(b) Similar data for SrFe$_{1.97}$Ni$_{0.03}$As$_2$.
The vertical blue and
red dashed lines in (a) and (b) mark the $T_N$ of the sample at zero and finite pressure, respectively.
(c) Similar data for BaFe$_{1.97}$Ni$_{0.03}$As$_2$, where the vertical green and blue dashed lines mark
$T_N$ and $T_s$, respectively, at zero pressure.
The open green and yellow squares
mark measurements of FWHM
under zero pressure (fresh) at the
$(4,0,0)$ and $(0,4,0)$ Bragg peaks, respectively.  The pink solid circles are identical
measurements under $P\approx 10$ MPa uniaxial pressure on $(4,0,0)$.  The
solid green diamonds are data after uniaxial pressure is released. The vertical red dashed line indicate the peak position of the FWHM under $P=10$ MPa.
(d) Temperature dependence of FWHM in $\Delta d/d$ for BaFe$_2$(As$_{0.7}$P$_{0.3}$)$_2$ at $P=0$ MPa (solid green circles),
BaFe$_{1.88}$Ni$_{0.12}$As$_2$ at $P=0$ (solid green diamonds) and 14 MPa (solid red circles).
}
\end{figure}

To determine the effect of uniaxial pressure on the tetragonal-to-orthorhombic phase transition
in iron pnictides, we carried out neutron Larmor diffraction experiments capable of measuring minor change of lattice
spacing $d=2\pi/|\textbf{Q}(H,K,L)|$ and its spread $\Delta d$ with a resolution
better than $10^{-5}$ in $\Delta d/d$ [inset in Fig. 3(a)] \cite{Bayrakci,supplementary}.
We focus on $(4,0,0)$ and $(0,4,0)$ nuclear
Bragg reflections corresponding to a $d$-spacing $d\approx a/4$, which we measured in
BaFe$_{2-x}$Ni$_{x}$As$_{2}$ ($x=0,0.03, 0.12$), SrFe$_{1.97}$Ni$_{0.03}$As$_2$,
and BaFe$_2$(As$_{0.7}$P$_{0.3}$)$_2$ both on freshly prepared samples (uniaxial pressure $P=0$) and
under uniaxial pressure ($P\approx 10,20$ MPa).
Figure 3 shows the temperature and pressure dependence of the $d$ spread for these samples. The $d$ spread are characterized by the FWHM (Full-Width-Half-Maximum) of the lattice spacing distribution $f(\Delta d/d)$, which is assumed to be Gaussian distribution \cite{supplementary}. The diamonds in Figure 3(a) show temperature dependence of the
FWHM for BaFe$_2$As$_2$ at zero pressure.  Similar to BaFe$_{1.97}$Ni$_{0.03}$As$_{2}$ \cite{HRMan}, temperature dependence of FWHM follows a Curie-Wiess form and peaks around the zero-pressure value of $T_N\approx T_s$.
Upon application of a uniaxial pressure $P\approx 20$ MPa, the magnitude of FWHM increases at all temperatures and now peaks at an enhanced $T_N=144$ K [Fig. 3(a)].

Figure 3(b) shows similar data for SrFe$_{1.97}$Ni$_{0.03}$As$_2$, where there are coupled strong first order structural and AF phase transitions at $T_N=T_s=175$ K \cite{RWHu}. Compared with BaFe$_2$As$_2$, where the AF phase transition is weakly first order and structural transition is second order \cite{mgkim}, the AF and structural transition induced changes in FWHM are much smaller and confined to temperatures close to $T_N\approx T_s$ in SrFe$_{1.97}$Ni$_{0.03}$As$_2$ [Fig. 3(b)].  Under a uniaxial pressure $P\approx 20$ MPa, however,  both the FWHM and $T_N$ increase dramatically with solid lines showing Curie-Wiess fits to the data.
For BaFe$_{1.97}$Ni$_{0.03}$As$_{2}$, application of a $P\approx 10$ MPa
uniaxial pressure transforms temperature dependence of the FWHM, which
forms a broad peak above the zero-pressure value of $T_s$. Upon releasing the uniaxial pressure [$P$ released, filled green diamonds in Fig. 3(c)], the system goes back to the original unpressured fresh state.

Figure 3(d) compares temperature dependence of the FWHM for electron overdoped
BaFe$_{1.88}$Ni$_{0.12}$As$_{2}$ and BaFe$_2$(As$_{0.7}$P$_{0.3}$)$_2$, where both
materials are in the paramagnetic
tetragonal state without static AF order.
The weak temperature dependence of FHWM in
these materials suggests that the large temperature dependence of  FWHM in AF ordered BaFe$_{2-x}$Ni$_{x}$As$_{2}$ ($x=0,0.03$) and SrFe$_{1.97}$Ni$_{0.03}$As$_2$
is due to a strong magnetoelastic coupling.  Although application of a
 $P\approx 14$ MPa uniaxial pressure on BaFe$_{1.88}$Ni$_{0.12}$As$_{2}$ increases
the absolute value of FWHM, it is still weakly temperature dependent [Fig. 3(d)].

To further demonstrate the impact of
uniaxial pressure on the tetragonal-to-orthorhombic structural
transition in BaFe$_{2-x}$Ni$_{x}$As$_{2}$ ($x=0,0.03, 0.12$) and SrFe$_{1.97}$Ni$_{0.03}$As$_2$,
we compare in Figure 4 temperature dependence of the lattice parameters
along the orthorhombic $a$ and $b$ axis directions under zero and finite uniaxial pressure.
We first discuss results for BaFe$_{2-x}$Ni$_{x}$As$_{2}$ with $x=0$ [Fig. 4(a), 4(b)]
and 0.03 [Fig. 4(c) and 4(d)].
At $P=0$, the lattice parameters have $a=b$ at temperatures above $T_s$ (tetragonal phase)
and decrease linearly with decreasing temperature
[open diamonds and hexagons in Fig. 4(a) and 4(c)].
Upon application of a uniaxial pressure, the system becomes orthorhombic at all temperatures and
the orthorhombic structural transition becomes a crossover [filled diamonds and hexagons in
Fig. 4(a) and 4(c)].
Figures 4(b) and 4(d) show temperature dependence of the lattice orthorhombicity
 $\delta=(a-b)/(a+b)$
at different uniaxial pressures for $x=0$, and 0.03, respectively. For unpressured fresh samples ($P=0$), and after the pressure has been released,
the tetragonal structure becomes orthorhombic below $T_s$ and
the AF order below $T_N$ further enhances the lattice orthorhombicity \cite{mgkim}.
Upon applying the uniaxial pressure $P\approx 10$, 15, and 20 MPa, the temperature dependence of the
lattice orthorhombicity becomes remarkably similar to that of the
$B_{2g}$ elastoresistance and nematic susceptibility
of BaFe$_{2-x}T_{x}$As$_{2}$ obtained from transport \cite{HHKuo2014,HHKuo2015} and
elastic shear modulus/ultrasound spectroscopy measurements \cite{Fernandes10,Yoshizawa,anna}, respectively.

\subsection{Theoretical Ginzburg--Landau analysis}
To understand the temperature dependence of the pressure-induced lattice orthorhombicity
described in Figs. 4(b) and 4(d), we consider the Ginzburg-Landau free energy formalism used in previous works \cite{chu12,anna}:
\begin{equation}
F[\varphi,\delta] = F_0 + \frac{a}{2}(T-T_0)\varphi^2 + \frac{\tilde{B}}{4}\varphi^4 + \frac{C_{66,0}}{2}\delta^2 - \lambda \delta\varphi - P\delta,
\label{eq.landau}
\end{equation}
where the electronic nematic order parameter $\varphi$ is
coupled linearly to the orthorhombic lattice distortion $\delta$.
It then follows that  (see appendix \cite{supplementary})
\begin{equation}
\delta = (\lambda \left\langle \varphi \right\rangle + P)/C_{66,0},
\label{eq.delta}
\end{equation}
where
$C_{66,0}$ is the bare elastic constant that has no strong
temperature dependence and $P$ is the conjugate uniaxial pressure (stress)
\cite{xyscience,supplementary,Fernandes10,Yoshizawa,anna}.
In the absence of the elasto-nematic coupling ($\lambda=0$), the nematic susceptibility $\chi_\varphi=1/[a(T-T_0)]$ is characterized by the Curie-Weiss temperature $T_0$.
Upon considering the coupling
between the nematic order parameter $\varphi$ and the structural lattice distortion $\delta$ (or equivalently, the elastic shear strain $\varepsilon_6$), the elastic susceptibility takes on the form \cite{chu12,anna}:
\begin{equation}
\frac{\ud\delta}{\ud P}=\frac{1}{C_{66,0}}\frac{T-T_0}{T-T_{s}^{CW}} \\
\label{eq.suscept}
\end{equation}
with the renormalized nematic transition temperature
$T_s^{CW}=T_0+{\lambda^2}/{(aC_{66,0})}$ that is increased compared to the bare Curie-Weiss temperature $T_0$.
The pressure-induced lattice distortions in Figs. 4(b) and 4(d) can be well described by
the Curie-Weiss functional form  \cite{supplementary}. Therefore, uniaxial pressure induced
orthorhombic lattice distortion and its temperature dependence
in undoped and underdoped BaFe$_{2-x}$Ni$_{x}$As$_{2}$
are directly associated with the nematic susceptibility \cite{HHKuo2015,anna}. Since the external uniaxial pressure explicitly breaks the tetragonal lattice symmetry, it turns the nematic transition at $T_s=T_s^{CW}$ into a crossover, as is clearly seen in Fig. 4.

\begin{figure*}[htbp]
\includegraphics[width=14cm]{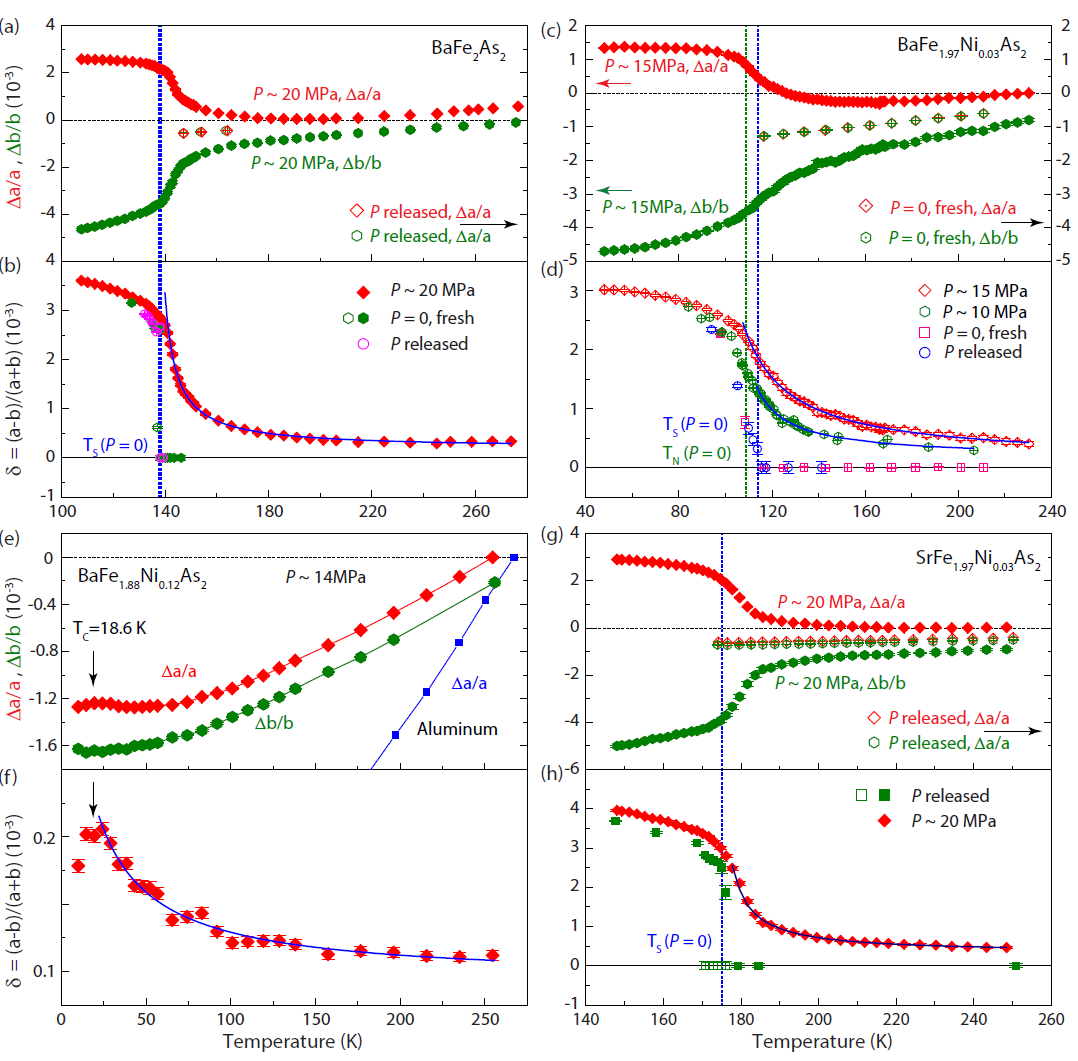}
\caption{ (Color online) Temperature dependence of the $a$ and $b$ lattice parameters and
orthorhombicity $\delta$
under different uniaxial pressure conditions ($P=0$ fresh, $\sim$10, $\sim$15, $\sim$20
and 0 released MPa) for BaFe$_{2-x}$Ni$_{x}$As$_{2}$ ($x=0,0.03,0.12$)
and SrFe$_{1.97}$Ni$_{0.03}$As$_2$.
(a) Temperature dependence of the $a$ and $b$
under $P=0$ and $20$ MPa uniaxial pressure for BaFe$_2$As$_2$.
(b) Temperature dependence of $\delta$
under different uniaxial pressure ($P=0$ fresh, $\sim$20,
and 0 released MPa).
The vertical blue dashed line marks the $T_N/T_s$.
(c) Temperature dependence of the lattice parameters $a$ and $b$
at $P=0$ and 15 MPa for BaFe$_{1.97}$Ni$_{0.03}$As$_{2}$.
(d) Temperature dependence of $\delta$
under different uniaxial pressure ($P=0$ fresh, $\sim$10, $\sim$15
and 0 released MPa).
The open red diamonds and green hexagons are
obtained by lattice thermal expansion measurements under uniaxial pressure.
The pink squares are measurements of an unpressured fresh sample,
while blue circles are obtained after releasing $P\approx 10$ MPa uniaxial pressure.
The blue circles and two pink squares below $T_s$
are from zero pressure Larmor diffraction measurements.
The pink squares above $T_s$ are obtained from thermal expansion measurements.
The vertical green and blue dashed lines in (c) and (d) mark the $T_N$ and $T_s$ of the sample at zero pressure, respectively.
(e) Temperature dependence of the $a$ and $b$ for BaFe$_{1.88}$Ni$_{0.12}$As$_{2}$. The lattice thermal expansion of aluminum
is plotted as a reference \cite{FCNix}.
The vertical arrow marks $T_c=18.6$ K and the solid lines are guides to the eye.
(f) Temperature dependence of the orthorhombic lattice distortions calculated from (c).
(g) Temperature dependence of the $a$ and $b$ lattice parameters for
SrFe$_{1.97}$Ni$_{0.03}$As$_{2}$ under $P=0$ fresh, 20 and 0 (released) MPa.
(h) Temperature dependence of $\delta$ for the same pressure condition.
The solid curves in (b), (d), (f), and (h)
are fits using a Curie-Weiss functional form \cite{supplementary}.
}
\end{figure*}

If the in-plane resistivity anisotropy in electron underdoped iron pnictides indeed
arises from the coupling of the uniaxial-pressure induced lattice distortion $\delta$ with the nematic susceptibility,
it would be interesting to determine the effect of similar uniaxial pressure on the electron overdoped sample, where
the resistivity anisotropy is known to be much weaker \cite{fisher}.
Figures 4(e) and 4(f) summarize the outcome of the neutron Larmor
diffraction experiments on uniaxial pressured BaFe$_{1.88}$Ni$_{0.12}$As$_{2}$, which is
tetragonal ($a=b$) and non-magnetic at all temperatures in zero pressure \cite{xylu13}.
Figure 4(e) shows temperature dependence of the lattice parameter
changes along the $a$-axis ($\Delta a/a$)
and $b$-axis ($\Delta b/b$) under a uniaxial pressure of $P\approx 14$ MPa.  For comparison, the thermal contraction of aluminum is also shown \cite{FCNix}.  Figure 4(f) shows the
temperature dependence of the orthorhombic lattice distortion
$\delta$, which reveals a clear anomaly at $T_c$ consistent
with ultrasonic spectroscopy measurements \cite{Fernandes10,Yoshizawa}.
While the applied uniaxial pressure
induces orthorhombic lattice distortion at 230 K,
the magnitude of the lattice distortion, $\delta\approx 1.1\times 10^{-4}$,
is about 5 times smaller than
that of BaFe$_2$As$_2$ and BaFe$_{1.97}$Ni$_{0.03}$As$_{2}$ at 230 K.  On cooling to 20 K,
$\delta$ in BaFe$_{1.88}$Ni$_{0.12}$As$_{2}$ increases to $\sim 2\times 10^{-4}$, while
$\delta$ in BaFe$_2$As$_2$ and
BaFe$_{1.97}$Ni$_{0.03}$As$_{2}$ becomes $\sim 2.5\times 10^{-3}$
near $T_s$ [Fig. 4(b) and 4(d)],
an order of magnitude larger than that of the electron overdoped compound.

To understand how a uniaxial pressure affects
the first order nature of the structural and magnetic phase transitions in
SrFe$_{1.97}$Ni$_{0.03}$As$_2$, we compare in Fig. 4(g) and 4(h) temperature dependence
of the lattice parameters and orthorhombicity under the zero and finite uniaxial pressure.
At zero pressure, the first order nature of the structural transition is clearly seen
in hysteresis of temperature dependence of the lattice parameters and distortion [Fig. 4(g) and 4(h)].
Upon application of $P\approx20$~MPa uniaxial pressure, the lattice orthorhombicity no longer displays the first order transition at $T_s$, but instead becomes a crossover, similar to that observed in the undoped and underdoped BaFe$_{2-x}$Ni$_{x}$As$_{2}$ [see Figs.~4(b) and 4(d)].

Assuming that the application of the modest uniaxial pressure $P\approx 20$~MPa can be considered in the linear-response regime~\cite{HRMan}, we can estimate the elastic susceptibility from the finite difference $\ud \delta /\ud P \propto \Delta(\delta)$ \mbox{$= \delta(P=20\ {\rm MPa})-\delta(P=0)$} and compare it among the different compounds in the iron pnictide family.
Figure 1(c) compares temperature dependence of $\delta(P=20\ {\rm MPa})-\delta(P=0)$
 for BaFe$_{2-x}$Ni$_{x}$As$_{2}$ ($x=0,0.03, 0.12$)
and SrFe$_{1.97}$Ni$_{0.03}$As$_2$ normalized for $P=20\ {\rm MPa}$.
For AF ordered BaFe$_{2-x}$Ni$_{x}$As$_{2}$ ($x=0,0.03$)
and SrFe$_{1.97}$Ni$_{0.03}$As$_2$, the magnitudes of
the pressure-induced lattice orthorhombicity are similar in the paramagnetic phase and vanish rapidly upon entering into the AF ordered state.
Furthermore, the $\delta(P=20\ {\rm MPa})-\delta(P=0)$ decreases for the iron pnictides
with reduced $T_N$, and are much smaller for BaFe$_{1.88}$Ni$_{0.12}$As$_{2}$.

\section{Discussion}

It is well known that the effect of increasing electron-doping in BaFe$_{2-x}T_{x}$As$_{2}$ is to suppress the
static AF order and to eliminate the low-temperature lattice orthorhombicity \cite{CLester2009,SNandi,hqluo,xylu13}.
At zero pressure, BaFe$_2$As$_2$ first exhibits a second-order structural transition from the high-temperature
paramagnetic tetragonal phase to a paramagnetic orthorhombic phase at $T_s$, followed by a discontinuous further
orthorhombic structural distortion and weakly first order AF phase transition at $T_N$ ($T_N<T_s$) due to
magnetoelastic coupling \cite{mgkim}.
Upon Ni-doping in BaFe$_{2-x}$Ni$_{x}$As$_{2}$, the structural and magnetic phase transitions are gradually
separated and suppressed [Fig. 1(a)], and become second order in nature \cite{CLester2009,SNandi,hqluo,xylu13}.
Upon application of a uniaxial pressure,
the $C_4$ rotational symmetry of the tetragonal lattice is broken.
Since the tetragonal-to-orthorhombic symmetry of the underlying lattice can only be broken once,
$T_s$ will become a crossover regardless
the magnitude of the applied pressure, as our findings in Figs. 3 and 4 corroborate. The same conclusion holds for SrFe$_{1.97}$Ni$_{0.03}$As$_2$ where the first-order structural transition becomes a crossover [see Figs. 4(g) and 4(h)].
 Therefore, both BaFe$_{2-x}$Ni$_{x}$As$_{2}$ and SrFe$_{1.97}$Ni$_{0.03}$As$_2$ under uniaxial pressure can only exhibit
AF phase transition. We note that our measurements and theoretical Landau--Ginzburg analysis do not rely on the microscopic nature of the nematic order parameter $\varphi$. In particular, they apply equally well to the so-called Ising spin nematic scenario \cite{RMfernandes11,fernandes14,CCL,jphu,si} or the  orbital order interpretation of nematicity \cite{myi,cclee,kruger,lv,ccchen,valenzeula,wang15}. In fact, the ferro-orbital order $\varphi_{orb} = \langle n_{xz} - n_{yz}\rangle$ is always linearly coupled~\cite{fernandes14,wang15} to the Ising spin nematic order parameter $\varphi_{spin} = \langle\vS_i\cdot \vS_{i+\hat{x}} - \vS_i\cdot \vS_{i+\hat{y}}\rangle$, so the orbital order is generically present whenever $\varphi_{spin}\neq 0$, although there are theoretical indications that the converse is not always true. In other words, the orbital order can exist in the absence of static AF order~\cite{wang15}, as is known to be the case in FeSe \cite{bendele10,baek15,anna15}. In either case, the application of external uniaxial stress renders the nematic transition a crossover, so that the lattice distortion $\delta$ and consequently $\varphi$ are both finite above the zero-stress value of $T_s$. In this light, the electronic anisotropy seen in the magnetic torque \cite{Kasahara} and scanning tunneling microscopy \cite{Rosenthal} measurements above $T_s$  without explicit external uniaxial pressure is likely due to intrinsic local strain in these materials which breaks the $C_4$ rotational symmetry of the paramagnetic tetragonal phase.  Indeed, local strain-induced effect has recently been observed in free standing BaFe$_2$As$_2$ above $T_N$ and $T_s$ \cite{XRen}.

The key finding of the present work is that undoped BaFe$_2$As$_2$, as well as  BaFe$_{1.97}$Ni$_{0.03}$As$_{2}$ and SrFe$_{1.97}$Ni$_{0.03}$As$_2$ all exhibit similar magnitudes of
the pressure-induced lattice orthorhombicity [Fig. 1(c) and Fig. 4] and FWHM of $\Delta d/d$ near $T_N$ (Fig. 3).
This indicates that
these samples experience similar strain field under nominally similar applied uniaxial pressure,
thus suggesting that the doped Ni impurities do not play an important role in determining
the strain field inside the sample. Theoretically, the electronic anisotropy of the iron
pnictides is expected to couple linearly to the lattice orthorhombicity $\delta$ \cite{CCL,jphu,si,fernandes14}, as captured by the effective Landau free energy in Eq.~(\ref{eq.landau}).
The Curie-Weiss like temperature dependence of the uniaxial pressure induced lattice distortion [Fig. 1(c)] is consistent with the temperature dependence of the nematic susceptibility $\ud\delta/\ud P$ in Eq.~(3) and agrees with the results of Young's modulus measurements \cite{anna}.
This gives us confidence that in the effective Landau description \cite{supplementary}, the uniaxial pressure-induced lattice distortion
$\delta$ has a component proportional to the electronic nematic order parameter $\varphi$ via Eq.~(2),
where one expects $\delta \propto \varphi$ in zero pressure ($P=0$).
Since $\delta$ has similar magnitude in BaFe$_2$As$_2$, BaFe$_{1.97}$Ni$_{0.03}$As$_{2}$, and SrFe$_{1.97}$Ni$_{0.03}$As$_2$ (see Fig. 4), one would also expect comparable values of $\varphi$ in all three compounds. So if one uses the resistivity anisotropy $\Delta\rho = (\rho_b-\rho_a)/(\rho_b+\rho_a)$ as a proxy for the nematic order parameter, as has been widely used in the literature~\cite{chu10,chu12,HHKuo2014,HHKuo2015}, how does one then explain the resistivity anisotropy differences in BaFe$_{2-x}T_{x}$As$_{2}$ \cite{fisher}
and a much smaller resistivity anisotropy above $T_N$ [Fig. 1(b)] in SrFe$_{2-x}T_{x}$As$_{2}$ family of materials \cite{SDDas}?
The bare value of the elastic shear modulus $C_{66,0}$ that enters Eq.~(\ref{eq.delta}) has no strong
temperature dependence \cite{anna} and from the Curie-Weiss fits of the nematic susceptibility to Eq.~(\ref{eq.suscept}), we find it to be roughly the same in all three compounds, $C_{66,0}\approx 50$~GPa \cite{supplementary}. The only remaining unknown variable is the elasto-nematic coupling constant $\lambda$, which could be material-dependent but not temperature-dependent \cite{chu12,anna}. It is thus very challenging to explain the qualitatively different temperature dependence of the resistivity anisotropy in BaFe$_{1.97}$Ni$_{0.03}$As$_{2}$ [monotonic, blue diamonds in Fig. 1(b)] from that in BaFe$_2$As$_2$ and in SrFe$_{1.97}$Ni$_{0.03}$As$_2$ [both non-monotonic, with a maximum at or just below $T_s$].
One possible explanation for the non-monotonic temperature dependence of the resistivity anisotropy, recently proposed in the context of FeSe \cite{Tanatar15}, is to assume a temperature dependent coefficient of proportionality between $\Delta\rho$ and $\varphi$:
\begin{equation}
\Delta\rho(T) = \Upsilon(T) \varphi(T),
\label{eq.rho}
\end{equation}
such that $\Upsilon(T)$ tends to zero as $T\to0$, whereas $\varphi(T)$ is expected to increase  monotonically below $T_s$ as the temperature is lowered (consider for instance the mean-field result $\varphi(T)\propto\sqrt{T_s-T}$ for the second order phase transition).

Even with the introduction of $\Upsilon(T)$ in Eq.~(\ref{eq.rho}), which has a meaning of the temperature-dependent scattering function, it is extremely difficult to explain the much lower value of $\Delta\rho$ in SrFe$_{1.97}$Ni$_{0.03}$As$_2$ compared to BaFe$_2$As$_2$ and BaFe$_{1.97}$Ni$_{0.03}$As$_{2}$. In fact, from the Curie-Weiss fits of the susceptibility data, we estimate the elasto-nematic coupling constant $\lambda$ to be a factor of $\sim 5$ smaller in SrFe$_{1.97}$Ni$_{0.03}$As$_2$ compared to BaFe$_{1.97}$Ni$_{0.03}$As$_{2}$~\cite{supplementary}. Given the comparable magnitudes of $\delta$ between the two compounds [see Fig. 4(d) and 4(h)], one would then expect  the nematic order parameter $\varphi$ to be a factor of $\sim 5$ greater in SrFe$_{1.97}$Ni$_{0.03}$As$_2$, to ensure that the left-hand side of Eq.~(2) remains of the same magnitude. And yet the resistivity anisotropy $\Delta\rho \propto\varphi$ paints a diametrically opposite picture, being much smaller in SrFe$_{1.97}$Ni$_{0.03}$As$_2$.

We propose that a likely resolution of this dilemma lies in the nature of the magnetic phase transition which we have so far neglected in our analysis. Indeed, it is well established that structural and magnetic phase transitions in SrFe$_{2-x}T_{x}$As$_{2}$ are coupled first order transitions that decrease with increasing $x$ before vanishing near optimal superconductivity \cite{RWHu}, while electron-doped BaFe$_2$As$_2$ has second order magnetic and
structural phase transitions \cite{CLester2009,SNandi,hqluo,xylu13}.
Although application of a uniaxial pressure renders the structural transition a crossover,
the first order nature of the magnetic transition means a vanishing
critical regime with suppressed low-energy
spin fluctuations at temperatures near $T_N$, compared with those of BaFe$_{2-x}T_{x}$As$_{2}$ where the AF phase transition is
second order. One expects the scattering of electrons on the magnetic fluctuations, and hence the resistivity, to therefore be smaller in the vicinity of the first-order magnetic transition, as is the case in  SrFe$_{2-x}T_{x}$As$_{2}$. We thus conclude that the vanishing resistivity anisotropy above $T_N$ in the uniaxial pressure detwinned SrFe$_{2-x}T_{x}$As$_{2}$ (compared with those of BaFe$_{2-x}T_{x}$As$_{2}$) is likely
rooted in the first order nature of the AF phase transition.  This is also consistent with the increased paramagnetic resistivity anisotropy on moving from
BaFe$_2$As$_2$ to BaFe$_{1.96}$Co$_{0.04}$As$_{2}$ \cite{ishida}, where the magnetic transition changes from weakly first order
to second order \cite{mgkim,CLester2009,SNandi}.  Similarly,
the lack of large resistivity anisotropy in the paramagnetic state of
uniaxial pressured Ba$_{1-x}$K$_x$Fe$_2$As$_2$ \cite{JJYing}, Ba$_{1-x}$Na$_x$Fe$_2$As$_2$ \cite{JQMa},
and Ca$_{1-x}$La$_x$Fe$_2$As$_2$ \cite{JJYing12} is likely due to
the first order nature of the paramagnetic to AF phase transition in these materials.
The phenomenological Landau theory can be extended to include the coupling between nematicity $\varphi$ and the magnetic order parameter~\cite{anna, supplementary}, and our theoretical analysis shows \cite{supplementary} that the resulting uniaxial pressure-induced lattice distortion $\delta(P)-\delta(0)$ reproduces semi-quantitatively the experimental findings in Fig.~1(c).

We conclude that
the in-plane resistivity anisotropy found in the paramagnetic state of iron pnictides depends sensitively on the nature of the magnetic phase transition and
a strong elasto-nematic coupling between the uniaxial pressure induced lattice distortion and the
electronic nematic susceptibility. We caution that while the resistivity anisotropy $\Delta\rho$ and its dependence on the shear strain can be successfully used to extract the quantity proportional to the nematic susceptibility \cite{chu12}, care should be taken when equating $\Delta\rho$ with the nematic order parameter itself. In particular, the non-monotonic temperature dependence of $\Delta\rho$ and its sensitivity to the nature of the magnetic phase transition remain relatively little explored and deserve further experimental and theoretical studies.

\section{ACKNOWLEDGEMENTS}

We thank A. E. B$\rm \ddot{o}$hmer, J. H. Chu, Jiangping Hu, and Qimiao Si for helpful discussions.
The neutron scattering work at Rice is supported by the
U.S. NSF-DMR-1308603, NSF-DMR-1362219, and DMR-1436006 (P.D.).
A.H.N. is supported by the U.S. NSF CAREER award (DMR-1350237).
This work is also supported by
the Robert A. Welch Foundation Grant Nos. C-1839 (P.D.) and C-1818 (A.H.N.).
The work at the Institute of Physics, Chinese Academy of
Sciences is supported by Ministry of Science and Technology
of China (973 project: 2012CB821400 and 2011CBA00110),
National Natural Science Foundation of China Projects 11374011 and 91221303
and The Strategic Priority Research Program (B) of the Chinese Academy of Sciences
Grant No. XDB07020300.

\section{APPENDIX}

\subsection{Sample Information}
The iron pnictide single crystals used in present study were prepared by self-flux method \cite{ycchen}. The samples have been characterized by resistivity, magnetization, and neutron scattering measurements. Figures 5 and 6 show the basic characterizations of BaFe$_{1.97}$Ni$_{0.03}$As$_{2}$ and SrFe$_{1.97}$Ni$_{0.03}$As$_{2}$ samples, respectively. The basic characteristics of the BaFe$_{2}$As$_{2}$ and BaFe$_{1.88}$Ni$_{0.12}$As$_{2}$ samples can be
found elsewhere \cite{ycchen,xyscience}.

\begin{figure}[htbp]
\includegraphics[width=7cm]{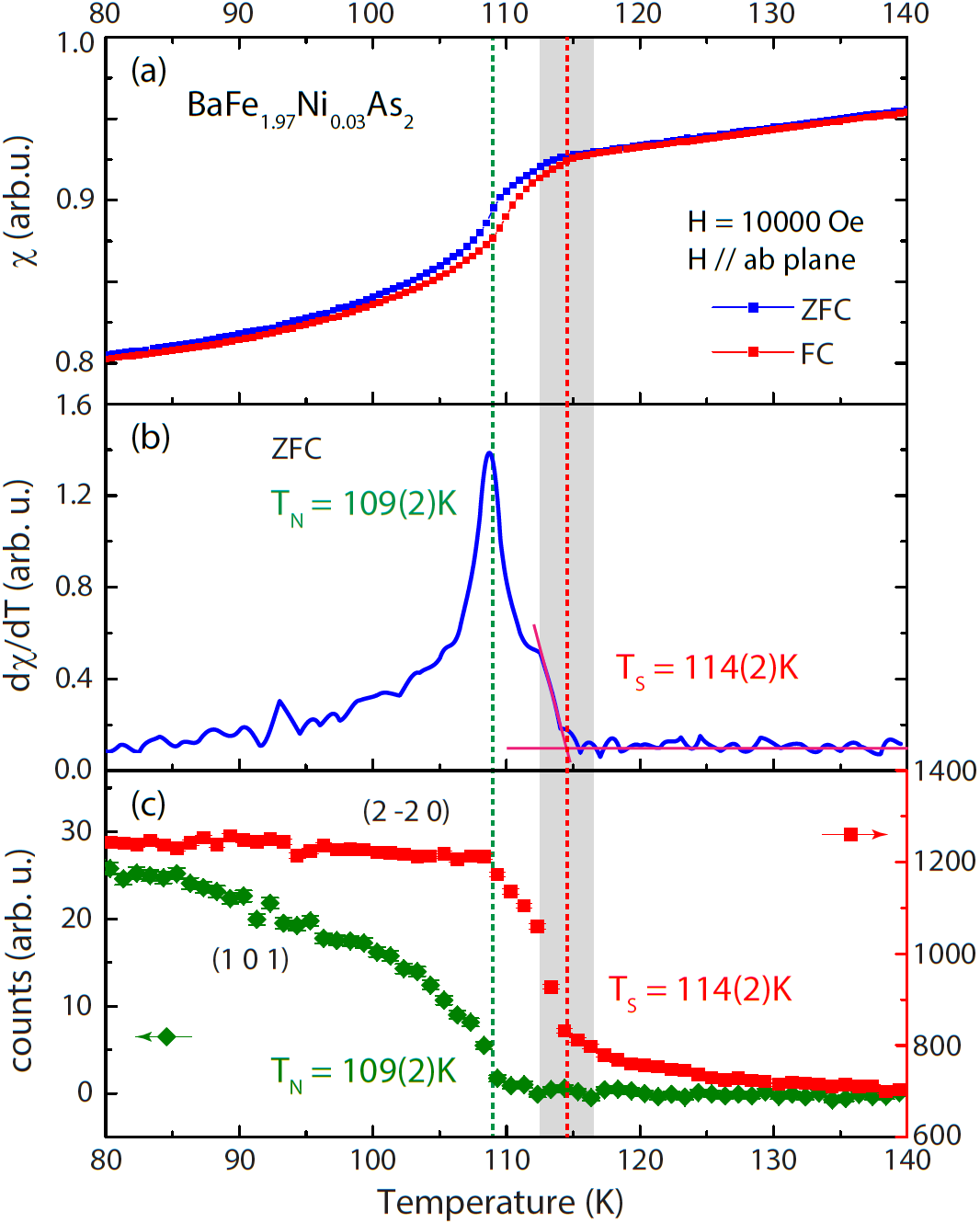}
\caption{ (Color online) (a) Temperature dependence of the magnetic susceptibility under ZFC and FC cases
for BaFe$_{1.97}$Ni$_{0.03}$As$_{2}$.
(b) Temperature derivative of the magnetic susceptibility, showing clearly the structural
and magnetic phase transitions.
 (c) Temperature dependence of the ${\bf Q}=(1,0,1)$ magnetic (green diamonds) and $(2,-2,0)$ nuclear (red squares)
Bragg peak intensity \cite{xyscience}. The extinction release of the $(2, -2, 0)$ Bragg reflection
is sensitive to the change of structural distortion and used to determine the $T_s$.
Combining the results in (a)-(c), the $T_N$ and $T_s$ are determined as shown in the green and red vertical
dashed lines, respectively.}
\label{fig:5}
\end{figure}

Figure 5(a) shows temperature dependence of the zero field cooled (ZFC) and
field cooled (FC) magnetic susceptibility $\chi$. Figure 5(b) is temperature derivative of $\chi$, $d\chi/d T$.
Figure 5(c) shows temperature dependence of the magnetic
$(1,0,1)$ Bragg peak (green diamonds) and $(2,-2,0)$
nuclear Bragg intensity (red squares).
These results establish $T_N$ (green dashed line) and $T_s$ (red dashed line) of BaFe$_{1.97}$Ni$_{0.03}$As$_{2}$.
The $T_N$ was determined as $109\pm 2$ K from magnetic order parameter of the $(1,0,1)$ magnetic Bragg peak [Fig. 5(c)] and
temperature dependent magnetization measurements [Fig. 5(a)]. The structural transition temperature
$T_s$ is estimated from a feature shown in the magnetization, [Fig. 5(b)],
and the neutron extinction release of the $(2, -2, 0)$ nuclear
Bragg peak intensity [Fig. 5(c)] \cite{xyscience,ninithesis,extinction}.

\begin{figure}[htbp]
\includegraphics[width=7cm]{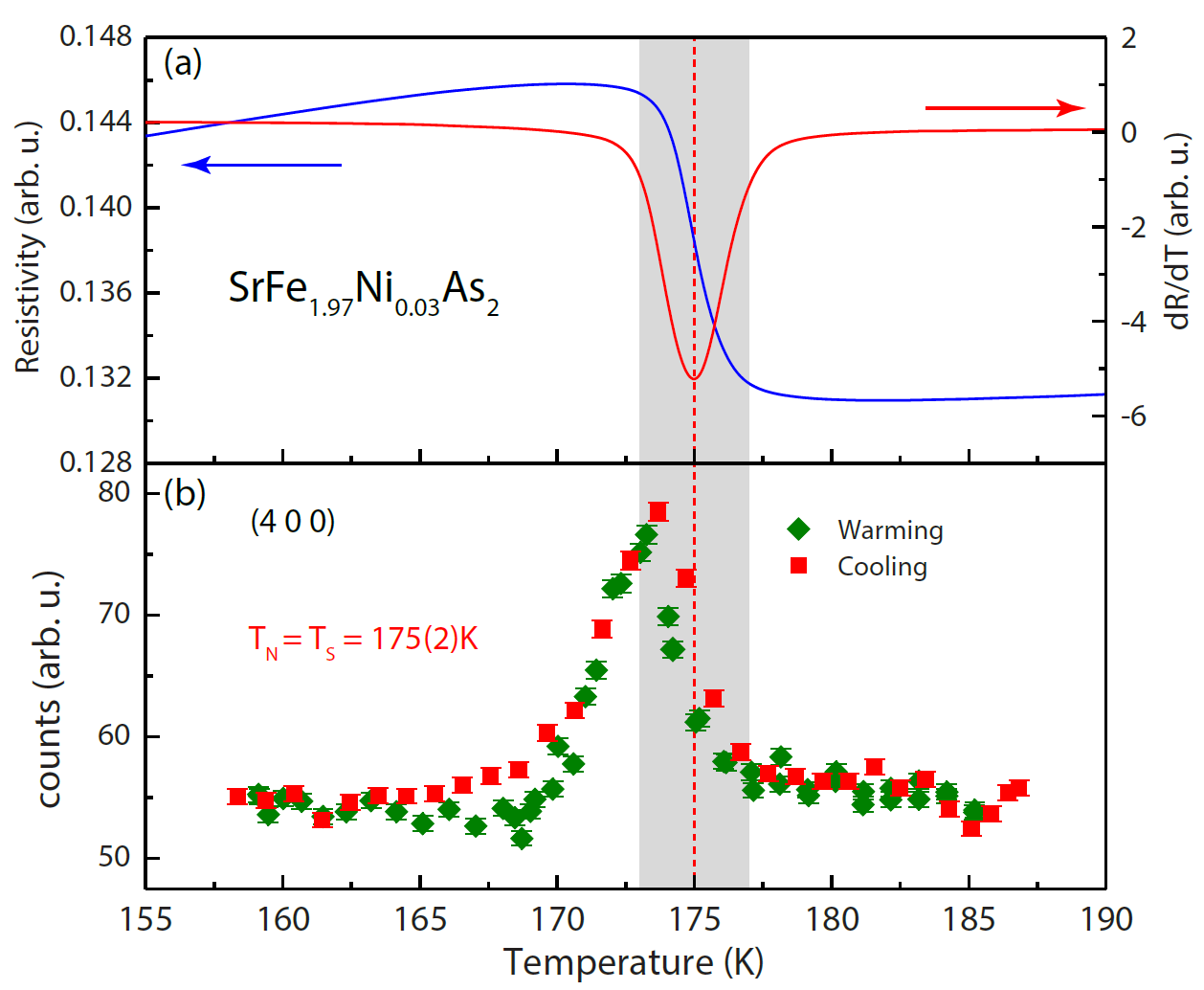}
\caption{ (Color online) (a) Temperature dependent resistivity and its temperature derivative for SrFe$_{1.97}$Ni$_{0.03}$As$_{2}$.
 (b) Temperature dependence of the $(4,0,0)$ nuclear Bragg peak intensity measured for warming and cooling. The dramatic increase of the $(4, 0, 0)$ Bragg reflection signals the change of structural distortion near $T_s$ \cite{xyscience}.}
\label{fig:6}
\end{figure}

Figure 6(a) shows temperature dependent resistivity $R$ and its derivative $dR/d T$ for SrFe$_{1.97}$Ni$_{0.03}$As$_{2}$.
The dip of the $dR/d T$ at $T=175$ K indicates the concomitant structural and magnetic transition, different from the two features for the separated $T_N$ and $T_s$ in BaFe$_{1.97}$Ni$_{0.03}$As$_{2}$. Figure 6(b) is temperature dependence of the $(4, 0, 0)$ nuclear Bragg peak intensity. The dramatic increase of the peak intensity also signals the structural transition. The observed intensity hysteresis is consistent with first order nature of the structural and
magnetic phase transition.

Large single crystals were selected and cut into rectangular shapes along the orthorhombic $[1,0,0]$ and $[0,1,0]$ directions
by a high precession wire saw. The well-cut samples were placed inside a uniaxial pressure device with
$b$ axis being the direction of the applied pressure \cite{xyscience}.
The applied uniaxial pressures for the the samples range from $P \sim 10$ MPa to $P \sim 20$ MPa, as described before.

In order to measure temperature and pressure dependence of orthorhombic lattice distortion ${\delta}=(a-b)/(a+b)$,
the samples were mounted in the $[H,K,0]$ scattering plane, where orthorhombic $(4,0,0)$ and $(0,4,0)$
Bragg reflections can be measured.
The effects of uniaxial pressure on tetragonal-to-orthorhombic structural transition and uniaxial-pressure induced lattice orthorhombicity can be probed
directly via measuring temperature and pressure dependence of the $(4,0,0)$ and $(0,4,0)$ reflections by neutron Larmor diffraction.
For magnetic measurements, the BaFe$_{1.97}$Ni$_{0.03}$As$_{2}$ sample was mounted in the $[1,1,2]\times [1,-1,0]$ scattering plane, where both the $(1,0,1)$ and $(0,1,1)$ magnetic Bragg peaks can be reached \cite{xyscience}.

\subsection{Neutron resonance spin echo measurements}

Neutron spin echo (NSE) technique has been demonstrated to be an effective method to measure the slow dynamics (quasielastic scattering) with an extremely high energy resolution ($\sim 1\ \mu$eV or even to $\sim 1$ neV) \cite{mezei}.
By combining triple axes spectrometer and neutron resonance spin echo (NRSE) technique, the TRISP spectrometer at
the Forschungsneutronenquelle Heinz Maier-Leibnitz (MLZ) is capable of measuring the lifetime of excitations with an energy resolution $\Delta E \sim 1\ \mu$eV in the range of about $1-200\ \mu$eV  \cite{trisp}.

\begin{figure}[htbp]
\includegraphics[width=8cm]{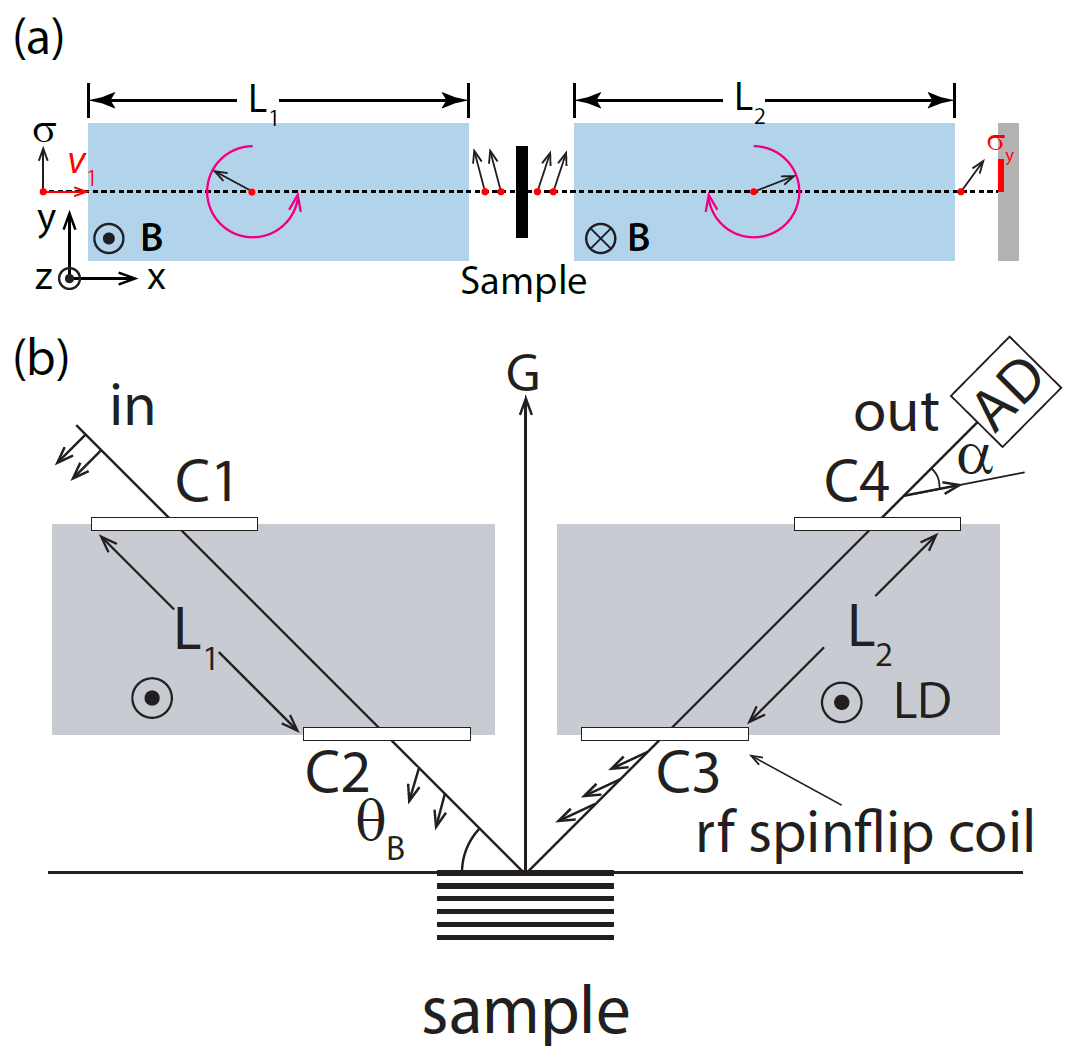}
\caption{(Color online) (a) Basic principles for neutron spin echo with schematics of the experimental setup.  The magnetic guide field ${\bf B}$
directions are clearly marked.
(b) Schematic diagram for the neutron Larmor diffraction measurements. For NRSE, the neutron precession direction in $L_1$ is opposite to that of $L_2$. In neutron Larmor diffraction, the neutron precession directions are same in $L_1$ and $L_2$.}
\label{fig:7}
\end{figure}

\begin{figure}[htbp]
\includegraphics[width=7cm]{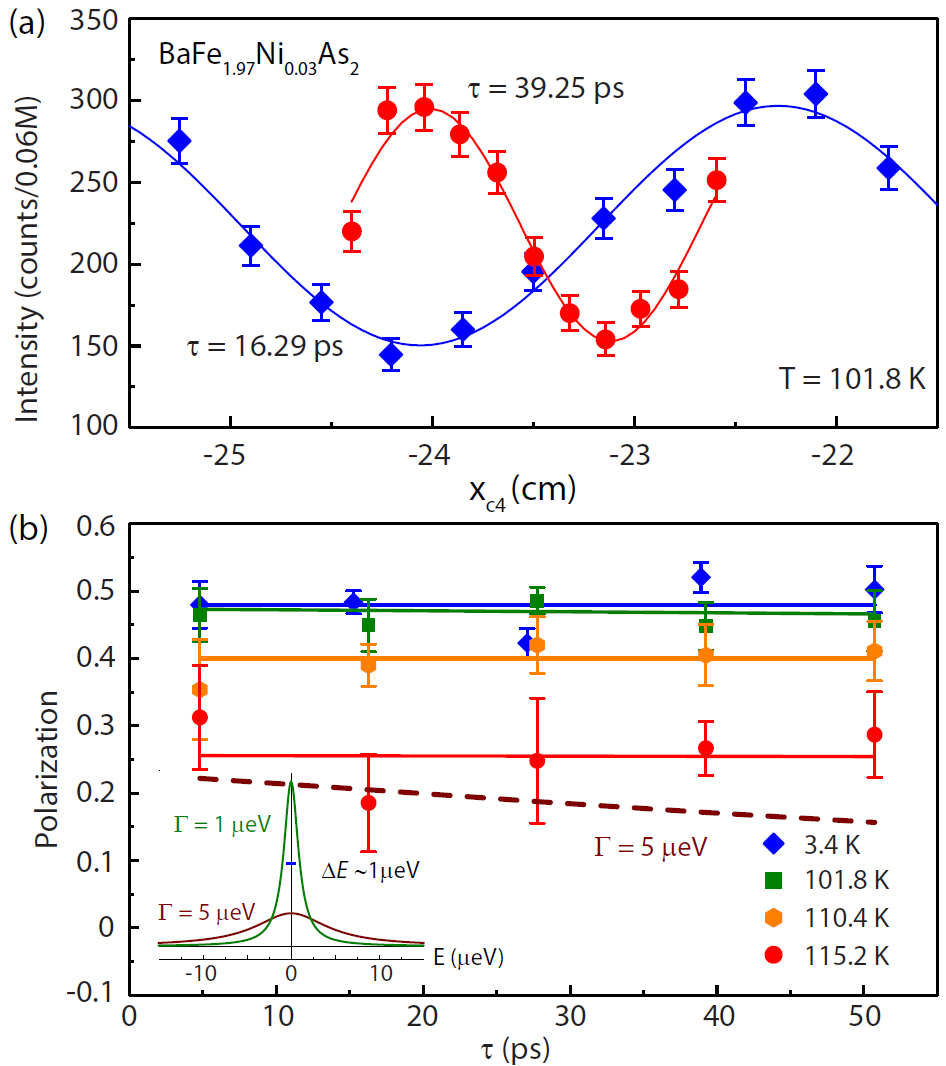}
\caption{ (Color online) (a) Measurements of the neutron polarization in one period of $\phi$ for $\tau=16.29\ ps$ and $39.25\ ps$. The solid lines are consine fittings of the data using eq. (8). (b) $P(\tau)$ for various temperatures. The solid lines are fittings by eq. (10). The brown dashed line shows a drawing of eq. (10) with $\Gamma = 5\ \mu$eV for comparison purpose. The corresponding Lorentzian $S(\omega)$ are shown as the solid green and brown lines in the inset.}
\label{fig:8}
\end{figure}

Compared with typical neutron scattering experiments where $S(\mathbf{Q},\omega)$ ($E=\hbar\omega$) is usually measured, neutron spin echo measures $I(\mathbf{Q}, \tau_{NSE})$ or $P(\mathbf{Q}, \tau_{NSE})$, where $P$ is the polarization of the scattered neutrons, which is the time Fourier transform of the $S(\mathbf{Q},\omega)$ and thus provides direct information of $S$(\textbf{Q},$\omega$) such as energy line-width (lifetime) and intensity \cite{Bayrakci,keller1}.

The basic principle of NSE can be understood in a simplified picture as shown in Figure 7(a). We assume neutrons polarized along the $y$ direction with a velocity $v_1$ enter the first arm of NSE spectrometer with a constant magnetic field $\mathbf{B}$ [Fig. 7(a)]. The precession angle in the first arm ($L_1$) is then $\phi_1 = \omega_L t=\gamma |B| L_1/v_1$, where $\gamma=2\mu_N/\hbar=2.916~k$Hz$/Gauss$ is the gyromagnetic ratio of neutron, $L_1$ is the length of the first neutron guide arm, and $t$ is the time for neutron to travel
through the first arm. After interactions with the sample, some neutrons are scattered into different energy with velocity $v_2$. In the second arm ($L_2$), the neutron spin will precess along the opposite direction, generating $-\phi_2 = -\omega_L L_2/v_2$. Assuming $L_1=L_2=L$ and $v_2=v_1+\delta v$, $\delta v<< v_1$, the net phase after passing through both
field regions will be \begin{math}\phi=\frac{\omega_L L}{v_1^2}\delta v\end{math}.
Since neutron energy transfer is \begin{math}\hbar\omega=\frac{1}{2}m(v_2^2-v_1^2)\approx m v_1\delta v\end{math}, the net phase can be written as
\begin{equation}
\phi=\left(\frac{\hbar\omega_L L}{mv_1^3}\right)\omega\equiv\omega\tau_{NSE}\label{eq:5}\\
\end{equation}
where $\tau_{NSE}$ is defined as
\begin{equation}
\tau_{NSE}=\left(\frac{\hbar\omega_L L}{m v_1^3}\right)=1.863 \times 10^{-16}B(gauss)L(cm)\lambda^3({\AA})\label{eq:6}\\
\end{equation}
Note $\tau_{NSE}$ is not a physical time but a quantity determined by specific parameters of the spectrometer, with the dimension of time.

The polarization along $y$ direction of the scattered neutrons can be analyzed and detected [Fig. 7(a)]. The average polarization $<\sigma_y>$ for neutrons with energy transfer $\hbar\omega$ is
\begin{equation}
<\sigma_y>=<\text{cos}~\phi>=\int d\omega S(\textbf{Q},\omega)\text{cos}~\omega\tau_{NSE}\text{.}\label{eq:7}\\
\end{equation}
 Thus $<\sigma_y>$ is the cosine Fourier transform of $S$(\textbf{Q},$\omega$) for $\omega$ and has been shown equal to the intermediate scattering function $I$(\textbf{Q},$\tau$). Therefore, the $\tau_{NSE}$ dependent polarization $P(\tau)$, that is, $I$(\textbf{Q},$\tau$), provide direct information about $S$(\textbf{Q},$\omega$) \cite{Bayrakci}.

 In the NRSE, the precession fields and spin flippers are replaced by four short bootstrap r-f spin flipper coils [C1-C4 in Fig. 7(b)], which can improve the energy resolution by a factor of 4 compared with the NSE with the same \textbf{B} and $L$. The neutrons only precess in bootstrap while keep their spin directions in $L_1$ and $L_2$. $L_2$ can be tuned by translating the flipper $C4$, by which the intensity with respect to the position of $C4$, $I(x_{c4})$, can be measured.
For a fixed $\tau$, the measured intensity can be described as
 \begin{equation}
I(x_{c4})=\frac{I_0}{2}\Big[1+P \text{cos}\big[\frac{2\pi}{\Delta x_{c4}}(x_{c4}-x_{c4,0})\big]\Big]\text{,}\label{eq:8}\\
\end{equation}
 where $P$ is the polarization, $I_0$ is the averaged intensity of the scattered beam, $\Delta x_{c4}$ is the period of the intensity modulation, and $x_{c4,0}$ is the reference position of $C4$.

The measurements of the $P(\tau)$ for BaFe$_{1.97}$Ni$_{0.03}$As$_{2}$ under $P\approx 15$ MPa are summarized in Figure 8. Figure 8(a) shows the intensity modulations for $\tau= 16.29~ps$ and $39.25~ps$ of \textbf{Q}=$(1,0,1)$ at $T = 102$ K. The polarizations are obtained through fitting the data by eq. (8). The fitted $P(\tau)$ for different temperatures are plotted in Figure 8(b). Assuming the possible broadening in energy of the magnetic reflections is caused by some slow dynamics (quasielastic scattering), the corresponding $S(\omega)$ can be described by a simple Lorentzian:
\begin{equation}
S(\omega)=\frac{1}{\pi}\frac{\Gamma}{(\omega-\omega_0)^2+\Gamma^2}\text{,}\label{eq:9}\\
\end{equation}
where $\Gamma$ ($\Gamma \ge 0$) (Half Width at Half Maximum) is the line-width of the quasielastic scattering ($\omega_0=0$).
Followed by eq. (9), the $P(\tau)$ should be fitted by the Fourier transform of eq. (9):
\begin{equation}
P(\tau)=P_0\text{exp}\big(-\frac{\Gamma\tau}{\hbar}\big)\text{,}\label{eq:10}\\
\end{equation}
All the $P(\tau)$ in our measurements can be well described by this exponential decay, as shown in Figure 8(b). The fitted energy line-widths $\Gamma$ are less than $1\ \mu$eV, meaning the signal are resolution limited at the measured temperatures. The comparison between $S(\omega)$ for resolution limited ($\Gamma \leq 1\ \mu$eV) and $\Gamma = 5\ \mu$eV is shown in the inset of Figure 8(b), as a reference. The temperature dependence of $\Gamma$ is shown in Figure 2(b) of the main text, which must have values greater than zero.  The large error bars for the values of
 $\Gamma$ near $T_N$ are due to experimental uncertainties of $P(\tau)$ in Fig. 8(b).

\subsection{Larmor diffraction measurements}

\begin{figure}[htbp]
\includegraphics[width=7.5cm]{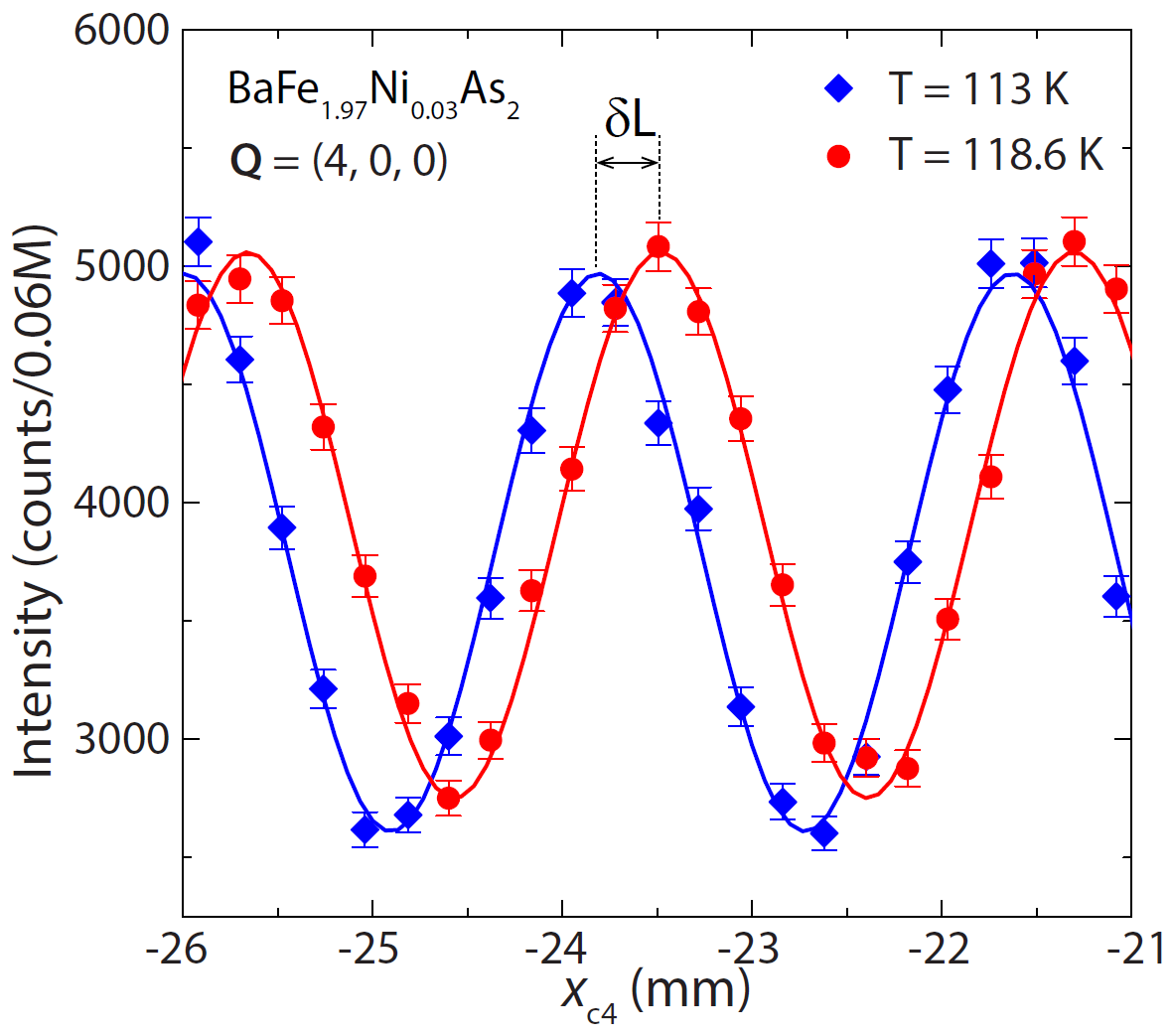}
\caption{ (Color online) Measurement of the $\Delta \phi_{tot}$ between $T=113$ K and $T=118.6$ K, by which the relative change of $d$ spacing can be tracked.}
\label{fig:9}
\end{figure}

\begin{figure}[htbp]
\includegraphics[width=7.5cm]{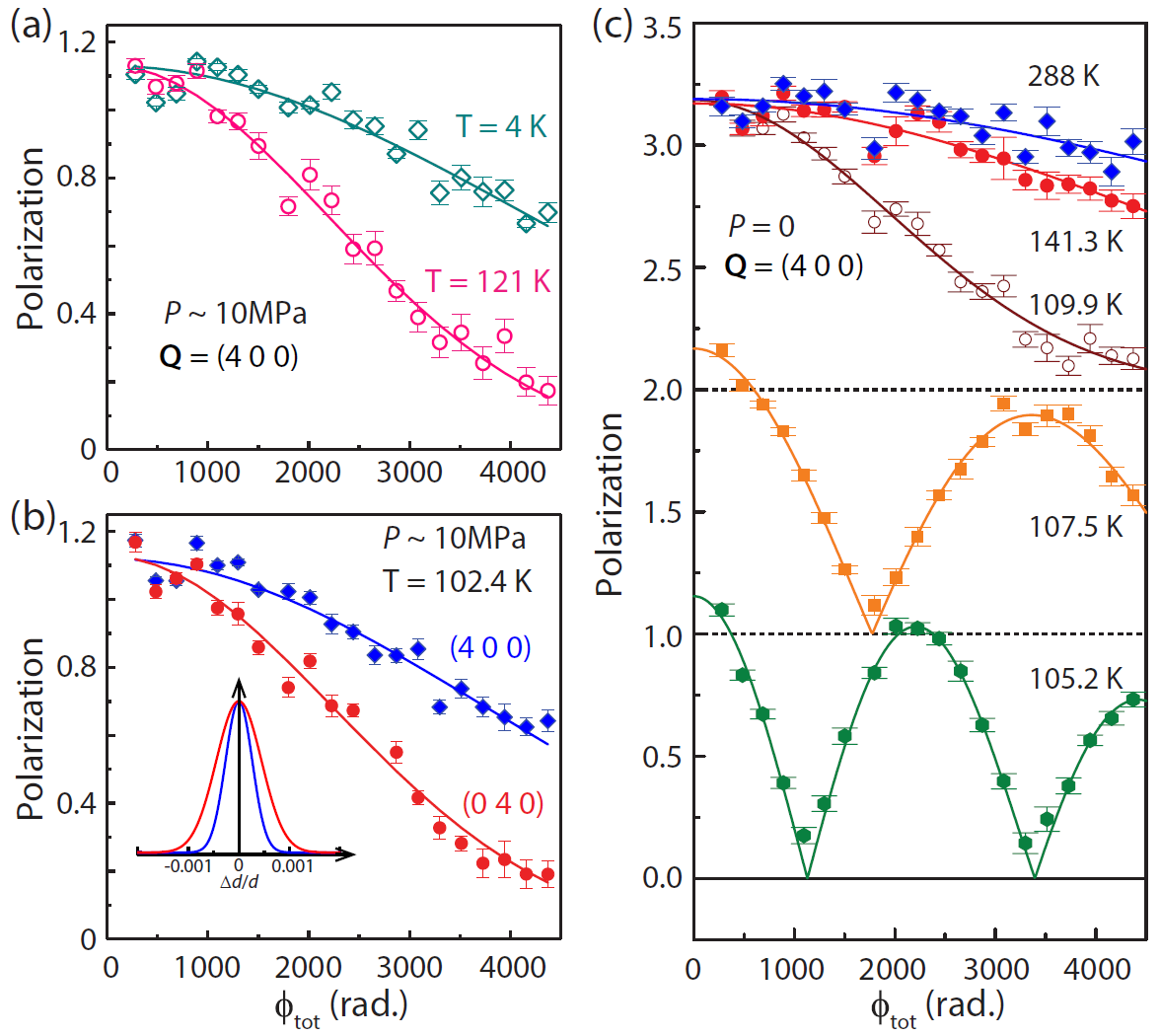}
\caption{(Color online) Larmor diffraction measurements of $P(\tau)$ for BaFe$_{1.97}$Ni$_{0.03}$As$_{2}$. (a) Comparison of $P(\phi_{tot})$ at $\textbf{Q}=(4,0,0)$ between $T=4$ K and $121$ K with $P\approx 10$ MPa. (b)
$P(\phi_{tot})$ for the $(4,0,0)$ and $(0,4,0)$ reflections measured at $T=102.4$
K with $P\approx 10$ MPa. The inset shows the Gaussian distribution of the $d$ spacing.
(c) Temperature dependence of the $P(\phi_{tot})$ for the $(4,0,0)$ peak at $P=0$.
$P(\phi_{tot})$ for the two $d$ spacings shows clear modulation
which can be fitted by eq. (14) (solid curves) due to the twinning caused by the tetragonal-to-orthorhombic
structural transition.}
\label{fig:10}
\end{figure}

\begin{figure}[htbp]
\includegraphics[width=7.5cm]{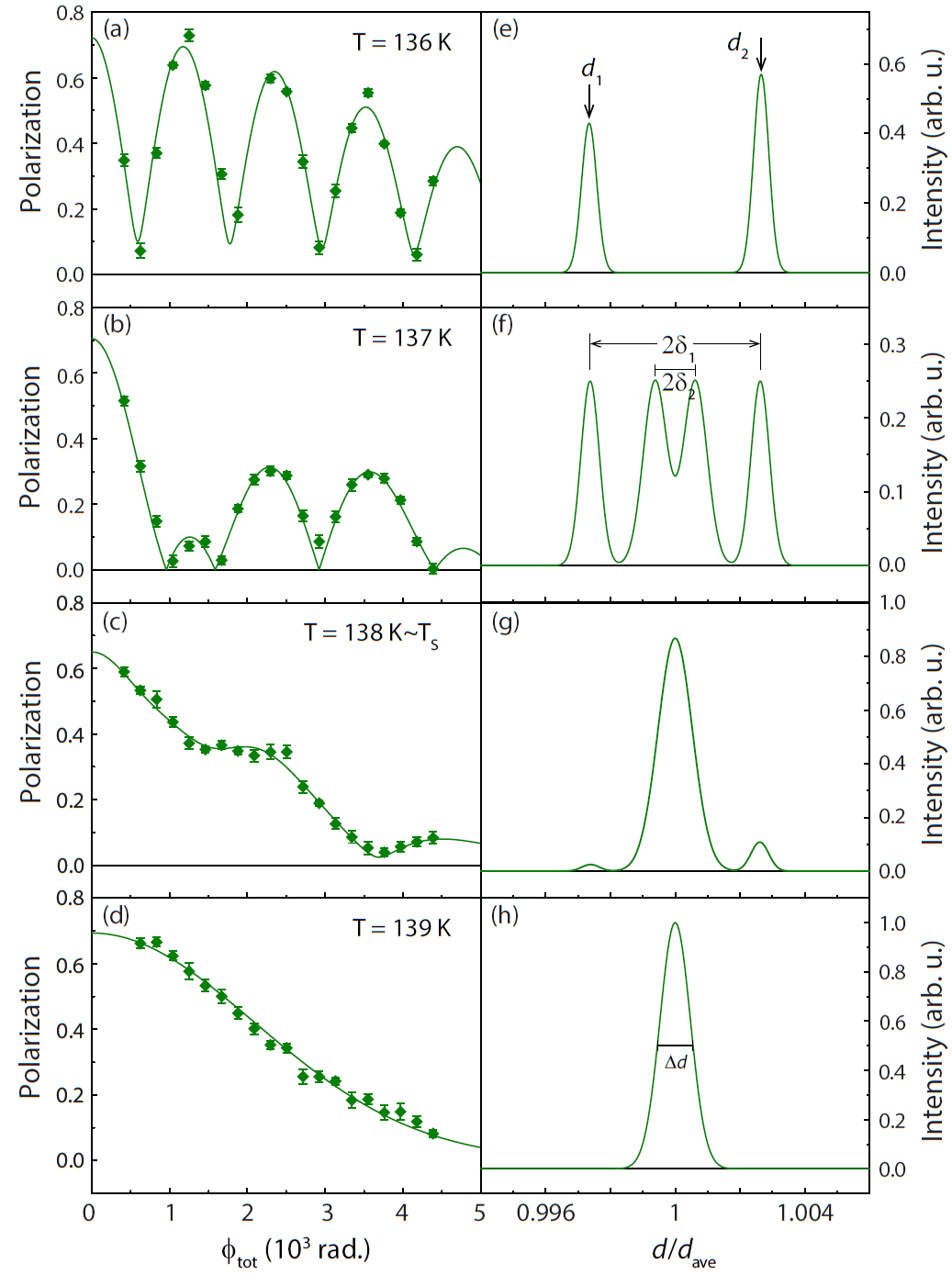}
\caption{ (Color online) (a)-(d) Precession phase ($\phi_{tot}$) dependent polarizations across the structural transition for BaFe$_{2}$As$_{2}$. The solid green curves are fits of the $P(\phi_{tot})$ by one or multi Gaussian $d$ spacing distribution models. (e)-(h) are the $d$ spacing distributions in $d$ space. The orthorhombic, coexisting two different orthorhombic and tetragonal phases can be determined for  $T=136$ to $139$ K, respectively.}
\label{fig:11}
\end{figure}

We now turn to the neutron Larmor diffraction measurements. Larmor diffraction is a neutron Larmor precession technique capable of measuring lattice spacing expansion and spread with a resolution better than $10^{-5}$ in term of $\Delta d/d$.  It is sensitive to minor change of lattice spacing $d$ but insensitive to sample mosaicity and not much affected by slight misalignment \cite{keller2}. The total precession phase ($\phi_{tot}$) dependent polarizations [$P(\phi_{tot})$] can be used to determine the $d$ spread and the splitting between multiple $d$ spacings with small differences, such as the peak splitting caused by the tetragonal-to-orthorhombic structural transitions in iron pnictides \cite{xyscience}.

Figure 7(b) is a schematic of Larmor diffraction. The spin flippers are tuned to be parallel with the diffracting planes and the neutron precession directions in $L_1$ and $L_2$ are the same. Assuming $L_1=L_2=L$, the total neutron precession phase is \begin{math}\phi_{tot} = 2\omega_L L/v\end{math}. From the Bragg law \begin{math}|\textbf{Q}|=|\textbf{G}|=2k_i\sin\theta_B, |\textbf{G}|=2\pi/d\end{math},
and the neutron velocity \begin{math}v=\hbar k_i/m\end{math}, the total Larmor phase $\phi_{tot}$ can be written as
\begin{equation}
\phi_{tot}=\frac{2\omega_LLm\sin\theta_B}{\pi \hbar}d\text{.}\label{eq:11}\\
\end{equation}
Consequently, the variation of the Larmor phase is proportional to the change of the $d$ spacing (caused by external or thermal effect), that is
\begin{equation}
\Delta\phi_{tot}=\phi_{tot}\frac{\Delta d}{d}\text{.}\label{eq:12}\\
\end{equation}
For $d$ change induced thermal expansion, the evolution of $P(\phi_{tot})$ at different temperatures and the relative change of the $\phi_{tot}$ ($\Delta \phi_{tot}$) can be obtained by fitting the intensity modulations $I(\phi_{tot,0}+\Delta \phi)$ using eq. (8), where the modulations are measured by scanning $x_{c4}$ near $x_{c4,0}$ [Fig. 7(b) and Fig. 9]. The $\Delta \phi_{tot}$ between different temperatures (or pressures/$\textbf{Q}$s) can be used to determine the evolution of the lattice spacings. To facilitate data analysis, $\Delta \phi_{tot}$ between two neighbouring conditions should be kept within $2\pi$. Figure 9 is an example of lattice thermal expansion at $\textbf{Q} = (4, 0, 0)$ of BaFe$_{1.97}$Ni$_{0.03}$As$_{2}$. The difference in $x_{c4}$ ($\delta L$) between $T = 113 $ K and $118.6 $ K, $\delta L$, can be converted to lattice expansions according to eq. (11, 12). In present measurements, $1$ mm is equivalent to $\sim 1\times 10^{-3}$ in $\Delta d/d$, with resolution $\sim 1\times 10^{-5}$. Note lattice expansion measurements is only valid for single $d$ spacing at one $\textbf{Q}$. Systems showing coexisting multiple $d$ spacings around the same $\textbf{Q}$ with small differences between them cannot be measured by this method.

Besides the lattice expansion measurements, the \begin{math}P(\phi_{tot})=<\text{cos}\Delta\phi(\phi_{tot})>\end{math} in a wide range of $\phi_{tot}$ has been demonstrated to be the Fourier transform of the lattice spacing distribution [$f(\Delta d/d)$] \cite{keller2,TKeller}. For a single-Gaussian distribution of $d$ with FWHM=$\varepsilon_{FW}$, the $P(\phi_{tot})$ can be derived as
\begin{equation}
P(\phi_{tot})=P_0\text{exp}\big(-\frac{\phi_{tot}^2 \varepsilon_{FW}^2}{16 \text{ln}2}\big)\text{,}\label{eq:13}\\
\end{equation}
where the FWHM represents the magnitude of the $d$ spread. It is usually expressed in term of $\Delta d/d$. The data shown in the Figure 3 of the main text are temperature dependence of the FWHM (lattice spacing spread). All of our $P(\phi_{tot})$ for single $d$ spacing are well described by this model, resulting in a Gaussian
distribution of the $d$ values. Figure 10 shows the $P(\phi_{tot})$ of BaFe$_{1.97}$Ni$_{0.03}$As$_{2}$ under $P \approx 10$ MPa and their fittings by eqs. (13),(14). A clear difference between $T=4$ K and $T=121$ K in Figure 10(a) indicates different FWHM of the $d$ spread. Figure 10(b) compares $P(\phi_{tot})$ for the $(4,0,0)$ and $(0,4,0)$ reflections at $T=102.4$ K. The corresponding FWHMs of the $d$ distributions are also shown as an inset. Their differences suggest that the $d$ spread along the pressure-applied orientation is much larger.

\begin{figure}[htbp]
\includegraphics[width=7.5cm]{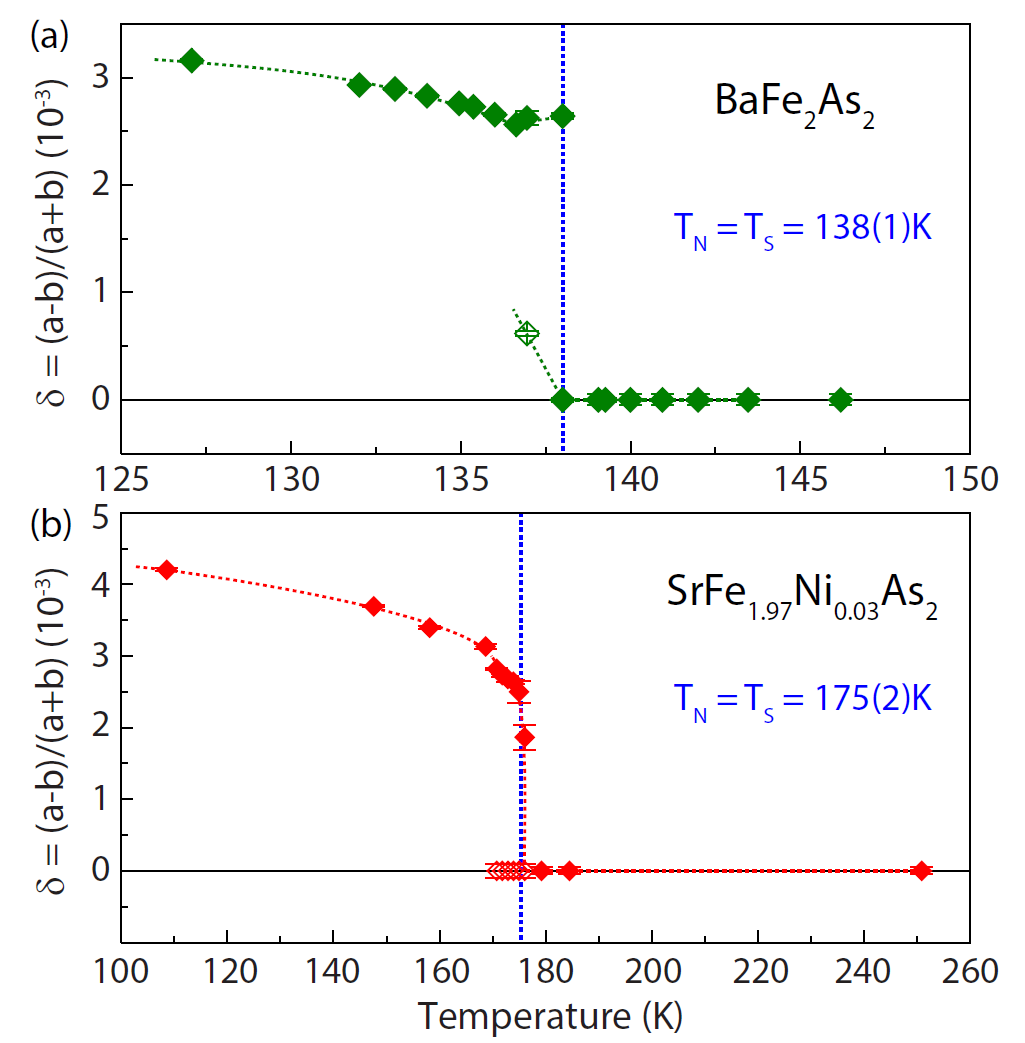}
\caption{(Color online) Temperature dependent orthorhombic lattice distortions for BaFe$_{2}$As$_2$ and SrFe$_{1.97}$Ni$_{0.03}$As$_{2}$. The open green diamond in (a) marks the temperature range showing four $d$ spacings. The open red diamonds in (b) show the persistence of the tetragonal phase into the orthorhombic phase, indicative of a first-order structural transition, consistent with previous reports. The vertical blue dashed lines mark the structural transitions.}
\label{fig:12}
\end{figure}

Figure 10(c) shows temperature dependence of $P(\phi_{tot})$ for an un-pressured BaFe$_{1.97}$Ni$_{0.03}$As$_{2}$ sample. Its evolution at high temperature ($T>109$ K) indicates the broadening of the $d$ spread. Below $109$ K, clear modulations are seen in $P(\phi_{tot})$. This is caused by the peak splitting of the orthorhombic $(4, 0, 0)$ and $(0, 4, 0)$ reflections in a twinned sample. For systems showing two or more coexisting $d$ spacings around some $\textbf{Q}$, their $d$ spacing distribution functions are superposition of multiple Gaussian distributions. In this case, the interference between different $d$ spacings will appear and can be used to identify the specific values and spread of the involved $d$ spacings.

For the peak splitting (two $d$ spacings) in BaFe$_{1.97}$Ni$_{0.03}$As$_{2}$, interference between scattered neutrons from $d_1$ and $d_2$ gives rise to the modulating polarization

\begin{equation}
P(\phi_{tot})=A\sqrt{a^2+(1-a)^2+2a(1-a)\text{cos}(\phi_{tot}\Delta\varepsilon)}\text{,}\label{eq:14}\\
\end{equation}
where
\begin{equation}
A=P_0\text{exp}\big(-\frac{\phi_{tot}^2 \varepsilon_{FW}^2}{16 \text{ln}2}\big)\text{,}\label{eq:15}\\
\end{equation}
here we assume both $d$ spacings have the same $\varepsilon_{FW}$. $a$ and $(1-a)$ denote the populations of the $d_1$ and $d_2$.
\begin{equation}
\Delta\varepsilon=\frac{d_1-d_2}{(d_1+d_2)/2}\text{,}\label{eq:16}\\
\end{equation}
is the lattice distortion. The definition of $\Delta\varepsilon$ is similar with the orthorhombic lattice distortion $\delta = (a-b)/(a+b)$ in iron pnictides, with $\Delta\varepsilon$ = $2\delta$ \cite{mgkim}. The $P(\phi_{tot})$ at $T=107.5$ K and $105.2$ K in Figure 10(c) are well described by eq. (14). The fitted lattice distortions and $d$ spreads are shown in Figure 3 and 4 of the main text. The resolution in determining $\Delta\varepsilon$ here depends on the range of $\phi_{tot}$ and the $d$ spread of the sample since the dips of the polarization is critical for fitting $\Delta\varepsilon$. The resolution of $\Delta\varepsilon$ for two $d$ spacings is $7\times10^{-4}$ in the present work. Thus the possible distortions at temperatures slightly lower than $T_s=114$ K [such as the $109.5$ K data shown in Figure 10(c)] in BaFe$_{1.97}$Ni$_{0.03}$As$_{2}$ cannot be distinguished from the broadening of the $d$ spread.

Figure {\ref{fig:11}} are $P(\phi_{tot})$ for temperatures across the structural transition of the BaFe$_{2}$As$_2$ sample. $P(\phi_{tot})$ in Figure {\ref{fig:11}}(a) is a beating pattern caused by interference between two $d$ spacings below $T_s$, similar with that shown in Figure 10(c). The corresponding $d$ spacing distributions are shown in Figure {\ref{fig:11}}(e). The orthorhombic distortions can be determined as $\delta=(d_2-d_1)/(d_2+d_1)$. Upon warming the sample to $T=137$ K, a temperature slightly lower than $T_s$, a more complicated pattern [Fig. {\ref{fig:11}}(b)] indicates the coexistence of four $d$ spacings [Fig. {\ref{fig:11}}(f)]. This is consistent with the coexisting orthorhombic antiferromagnetic ($\delta_1$) and orthorhombic paramagnetic ($\delta_2$) phases revealed by high resolution X-ray diffraction measurements \cite{mgkim}. The four-$d$ spacing model of $P(\tau)$ can be derived analytically (not shown here) and fit the data very well. Figure {\ref{fig:11}}(c) and (g) are results for $T \sim T_s$, where the $\delta_2$ is indistinguishable and only one broad $d$ spread can be fitted. Here, the orthorhombic antiferromagnetic phase ($\delta_1$) is about to disappear, suggesting this temperature is near $T_s$. For temperature higher than $T_s$ in Figure {\ref{fig:11}}(d) and (h), only one $d$ spacing is observed, indicating the system enters into the paramagnetic tetragonal phase. In Figure {\ref{fig:11}}(a,b), the magnitude of the lattice distortions determines the beating periods (overall line shape) and the relative populations of different $d$ spacings control whether the polarization can reach zero at dips. In the present study, the overall line shapes of all $P(\tau)$ are well fitted by specific multiple ($2-4$) $d$ spacing models [green curves in Figure {\ref{fig:11}}(a)-(d)], indicating that the lattice distortions are well determined.

The orthorhombic lattice distortions for BaFe$_{2}$As$_2$ and SrFe$_{1.97}$Ni$_{0.03}$As$_{2}$ obtained from Larmor diffraction measurements of $P(\phi_{tot})$ are shown in Figure {\ref{fig:12}}. These results are consistent with previous results measured by X-ray diffraction \cite{mgkim}.
 The error bars in Figures 3 and 4 of the main text are fitting errors of the raw data at different temperatures
 according to formulas discussed above.

\subsection{$d$ spread anisotropy between $a$ and $b$}

\begin{figure}[htbp]
\includegraphics[width=7.5cm]{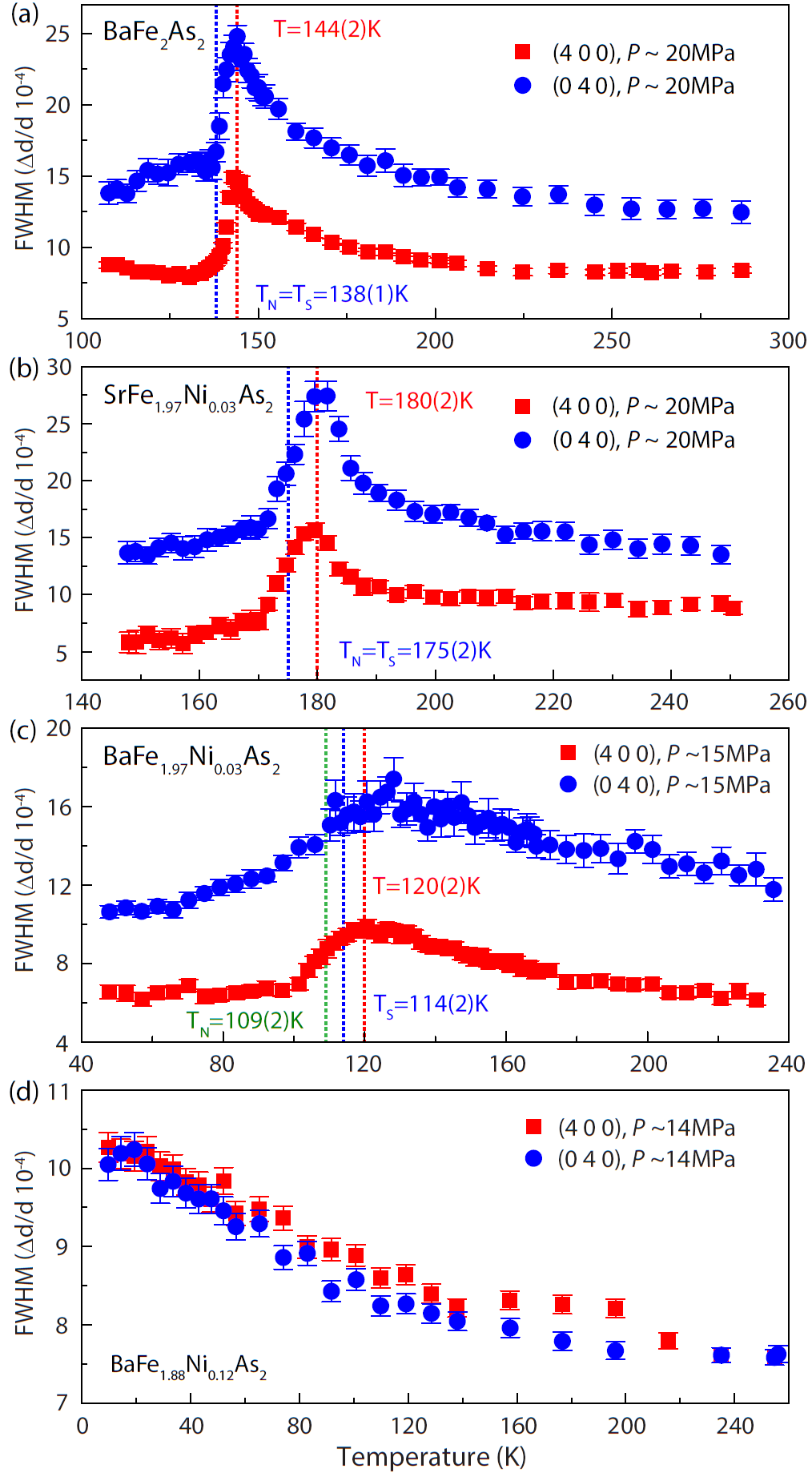}
\caption{(Color online) Temperature and doping dependent $d$ spread of BaFe$_{2-x}$Ni$_{x}$As$_{2}$ and SrFe$_{1.97}$Ni$_{0.03}$As$_{2}$ measured with finite uniaxial pressure. $T_N$ and $T_s$ are marked as blue (and green) vertical dashed lines. The red dashed line marks the temperature with the maximum of the FWHM.}
\label{fig:13}
\end{figure}

Another interesting discovery is the doping dependent $d$ spread anisotropy under uniaxial pressure. The samples shown in Figure 13 exhibit similar temperature dependence of the $d$ spread between $(4, 0, 0)$ and $(0, 4, 0)$, suggesting that the difference of $d$ spread between $a$ and $b$ is trivial. However, we note that the FWHM of $(0, 4, 0)$, along the uniaxial pressure direction, is much larger than $a$ in underdoped samples [Fig. 13(a)-13(c)].
This may be attributed to an inhomogeneous distribution of the pressure induced strain field.  However, we find very small differences in $d$ spread between $(4, 0, 0)$ and $(0, 4, 0)$ in the overdoped BaFe$_{1.88}$Ni$_{0.12}$As$_{2}$ [Fig. 13(d)], suggesting the $d$ spread anisotropy between $a$ and $b$ is non-trivial and may be associated with antiferromagnetic/structural instability or even nematic susceptibility in underdoped samples.

\subsection{The lattice distortions and Young's modulus}

The Young's modulus $Y$ along the $b$-axis ($\sim C_{66}$) can be estimated by \begin{math} Y= P/\delta \end{math}, where $\delta$ is pressure induced lattice distortion. At $\sim 250$K, the $Y$ for BaFe$_2$As$_2$, BaFe$_{1.97}$Ni$_{0.03}$As$_2$ and BaFe$_{1.88}$Ni$_{0.12}$As$_2$ estimated from our neutron Larmor diffraction experiments
are $\sim 50$ GPa, $\sim 50$ GPa and $\sim 100$ GPa, respectively. Compared with the shear modulus $C_{66}$ obtained by ultrasound spectroscopy \cite{Yoshizawa}, the estimated $Y$ for $x=0$ and $x=0.03$ are $\sim 30\%$ larger. These differences are mainly caused by the errors in our estimation of the applied pressure $P$ through measuring compressed spring distances and estimated spring constant \cite{xyscience}.
However, they will not affect temperature dependence of the pressure-induced
FHWM of $\Delta d/d$ and its comparison with other iron pnictides, thus will not alter the conclusions of our experiments.

\subsection{Landau theory and effect of magnetism on nematicity and strain}

In order to understand the distinct behavior of the observed lattice distortion in \SrFeNi compared to \BaFeNi [see Fig. 1(c) in the main text], we write down the Landau free energy  incorporating the electronic nematic order parameter $\varphi$, coupled magnetoelestically to the lattice distortion
$\delta \propto \eps_6$, as well as to the antiferromagnetic order parameter $M$:
\begin{equation}
F = F_0 + T_0 f[\varphi,\delta] + T_0 \tilde{f}[M,\varphi],
\end{equation}
where
\begin{equation}
f[\varphi,\delta] =  \frac{a}{2}\frac{T-T_0}{T_0}\varphi^2 + \frac{B}{4 T_0}\varphi^4 + \frac{C_{66,0}}{2T_0}\delta^2 - \frac{\lambda}{T_0} \delta\varphi - \frac{P\delta}{T_0}
\label{eq.phi-delta}
\end{equation}
Here we chose to normalize the free energy by the Curie-Weiss temperature $T_0$ associated with the quadratic $\varphi^2$ term (if $T_0$ is negative, it is replaced by $|T_0|$). Treating the electronic nematic order parameter $\varphi$ as a dimensionless variable, this has an advantage that all the coefficients in the free energy are dimensionless (here we choose, without loss of generality, \mbox{$a\!=\!B=1$}). The remaining Landau expansion parameters can be fixed from the experiment. Indeed, it is convenient to express the external uniaxial stress $P$ in terms of the dimensionless stress variable $\sigma = P/C_{66,0}$. Then, the last three terms in Eq.~(\ref{eq.phi-delta}) can be written as follows:
\begin{equation}
\frac{\lambda}{|T_0|}\left[  \left(\frac{C_{66,0}}{\lambda}\right)\left(\frac{\delta^2}{2} - \delta\sigma\right) - \delta \varphi \right]
\end{equation}
Minimizing the free energy with respect to $\delta$, we find
\begin{equation}
\delta = \frac{\lambda}{C_{66,0}}\varphi + \sigma.
\label{eq.delta}
\end{equation}
From the minimization with respect to $\varphi$, it is easy to obtain
\begin{equation}
\frac{\ud\varphi}{\ud\delta} = \frac{\lambda}{a(T-T_0) + 3B\phi^2}
\end{equation}
and now the shear modulus $C_{66} \equiv \ud P/\ud \delta = \ud^2 F/\ud \delta^2$ becomes
\begin{equation}
C_{66} = C_{66,0} - \lambda\frac{\ud \varphi}{\ud \delta} = C_{66,0} - \frac{\lambda^2}{a(T-T_0) + 3B\phi^2},
\label{eq.C66}
\end{equation}
in other words the elastic modulus gets renormalized from its bare value $C_{66,0}$ by virtue of the elasto-nematic coupling $\lambda$. Equivalently, it follows from the above equation that the inverse nematic susceptibility \mbox{$\chi_\varphi^{-1} \equiv \ud^2 F/\ud \varphi^2|_{\varphi\to0}$} also gets renormalized from its bare value $\chi_\phi^{-1} = a(T-T_0)$:
\begin{equation}
\tilde{\chi}_\varphi^{-1} = \chi_\phi^{-1} - \frac{\lambda^2}{C_{66,0}} = a(T-T_s^{CW}),
\label{eq.nematic-suscept}
\end{equation}
where $T_s^{CW} = T_0 + \frac{\lambda^2}{aC_{66,0}}$  is the renormalized Curie--Weiss temperature. One can now cast Eq.~(\ref{eq.C66}) above the transition temperature $T_s=T_s^{CW}$ into the form
\begin{equation}
\frac{\ud\delta}{\ud P}\equiv \frac{1}{C_{66}} =\frac{1}{C_{66,0}}\frac{T-T_0}{T-T_{s}^{CW}},  \qquad (T>T_s^{CW})
\label{eq.suscept}
\end{equation}
which is Eq. (3) in the main text. We now use this Eq.~(\ref{eq.suscept}) to fit the data for the pressure-induced distortion $\delta(P) - \delta(0) \approx P (\ud\delta/\ud P)$ [Fig. 1(c) in the main text], extracting the values of $T_0$ and $T_s^{CW}$ which we quote in Table \ref{table} for BaFe$_2$As$_2$ and \SrFeNi.

We now turn to the question of the strength of the elasto-nematic coupling constant $\lambda$.
The unknown dimensionless parameter $r=C_{66,0}/\lambda$ in Eq.~(\ref{eq.delta}) can be fixed from the ratio $\varphi/\delta$ in zero external stress ($\sigma=0$). Substituting the typical value of $\delta \sim 3\times 10^{-3}$ in \SrFeNi and \BaFeNi and choosing the value of the nematic order parameter $\varphi=1$ deep inside the nematic phase for convenience, we find $r \sim 300$. As for the value of $\tilde{\lambda} = \lambda/|T_0|$ itself, it can also be fixed from the experiment since $\lambda$ enters in Eq.~(\ref{eq.nematic-suscept}) to renormalize the Curie-Weiss temperature. Therefore,
one obtains \mbox{$\tilde{\lambda} = a \frac{(T_s^{CW} - T_0)}{|T_0|} r$}. Plugging in the values of $T_s^{CW}$ and $T_0$ from our fittings of the lattice distortions (Table \ref{table}), we obtain $\tilde{\lambda} \approx 80$ for \SrFeNi and $\tilde{\lambda} \approx 490$ for \BaFeAs, in other words the effective electron-lattice coupling is about $\sim\!6$ times weaker in \SrFeNi compared to the BaFe$_{2-x}T_x$As$_2$ compounds. For \BaFeNi, the quality of our data was insufficient to accurately determine the bare Curie--Weiss temperature $T_0$ (we were only able to determine $T_s^{CW}=88.5\pm1.0$~K). However, from the estimated  $T_s^{CW} - T_0 \approx 50$~K by the elastic measurements \cite{anna}, we can deduce the approximate value of the coupling constant $\tilde{\lambda}\approx 390$, similar in magnitude to undoped BaFe$_2$As$_2$.

\begin{table}[!h]
\caption[]{Curie--Weiss fitting parameters of the pressure-induced lattice distortions in the tetragonal state, see Fig. 1(c) in the main text.}
 \begin{tabular}{|p{2.6cm}|p{1.6cm}|p{1.6cm}|p{1.9cm}|}\hline
Sample     &$T_{s}^{CW}$(K)    &$T_0$(K)      &$T_s^{CW}-T_0$(K)   \\
\hline
BaFe$_2$As$_2$      &$134.9\pm0.3$    &$51.3\pm8.7$     &$83.5\pm8.7$ \\
\hline
SrFe$_{1.97}$Ni$_{0.03}$As$_2$       &$170.9\pm1$    &$135.7\pm6$   &$35.2\pm6.1$   \\
\hline
 \end{tabular}
\label{table}
\end{table}

We now turn our attention to the magneto-nematic coupling. On symmetry grounds, nematic order parameter must couple to $M^2$ (since magnetization breaks time-reversal symmetry, and $\varphi$ does not). This can be shown explicitly by considering the magnetization $\MM_{A,B}$ on the two sublattices composed of the next-nearest neighbor sites of the square lattice, in which case the nematic order parameter couples linearly to the product $(\MM_A\cdot \MM_B)$ [\onlinecite{jphu, Xu08, RMfernandes11, wang15}]. Note that this conclusion holds independently of whether the microscopic origin of nematicity is purely magnetic~\cite{jphu, Xu08, RMfernandes11} or orbital~\cite{Kruger09, cclee, lv, Chen09, wang15}. The magnetic phase transition itself may be intrinsically second order, as in BaFe$_{2-x}T_x$As$_2$ compounds, or first order, as in \SrFeNi, Ba$_{1-x}$(K,Na)$_x$Fe$_2$As$_2$, and Ca$_{1-x}$La$_x$Fe$_2$As$_2$. Below we consider both possibilities:
\begin{eqnarray}
\tilde{f}_1[M,\varphi]\!\! &=&\!\! \frac{u}{2}(T-T_{N,0})M^2 - \frac{v}{4}M^4 + \frac{w}{6}M^6 - \mu\varphi M^2 \phantom{xxx} \label{eq.first} \\
\tilde{f}_2[M,\varphi]\! \!&=&\!\! \frac{u}{2}(T-T_{N,0})M^2 + \frac{v}{4}M^4  - \mu\varphi M^2
\label{eq.second}
\end{eqnarray}
Since we are after the qualitative consequences of the magneto-nematic coupling, the precise values of the Landau coefficients are not essential (we take $u=v=w=1$ and $\mu=0.1$ for concreteness).

\begin{figure}[htb]
\includegraphics[width=0.44\textwidth]{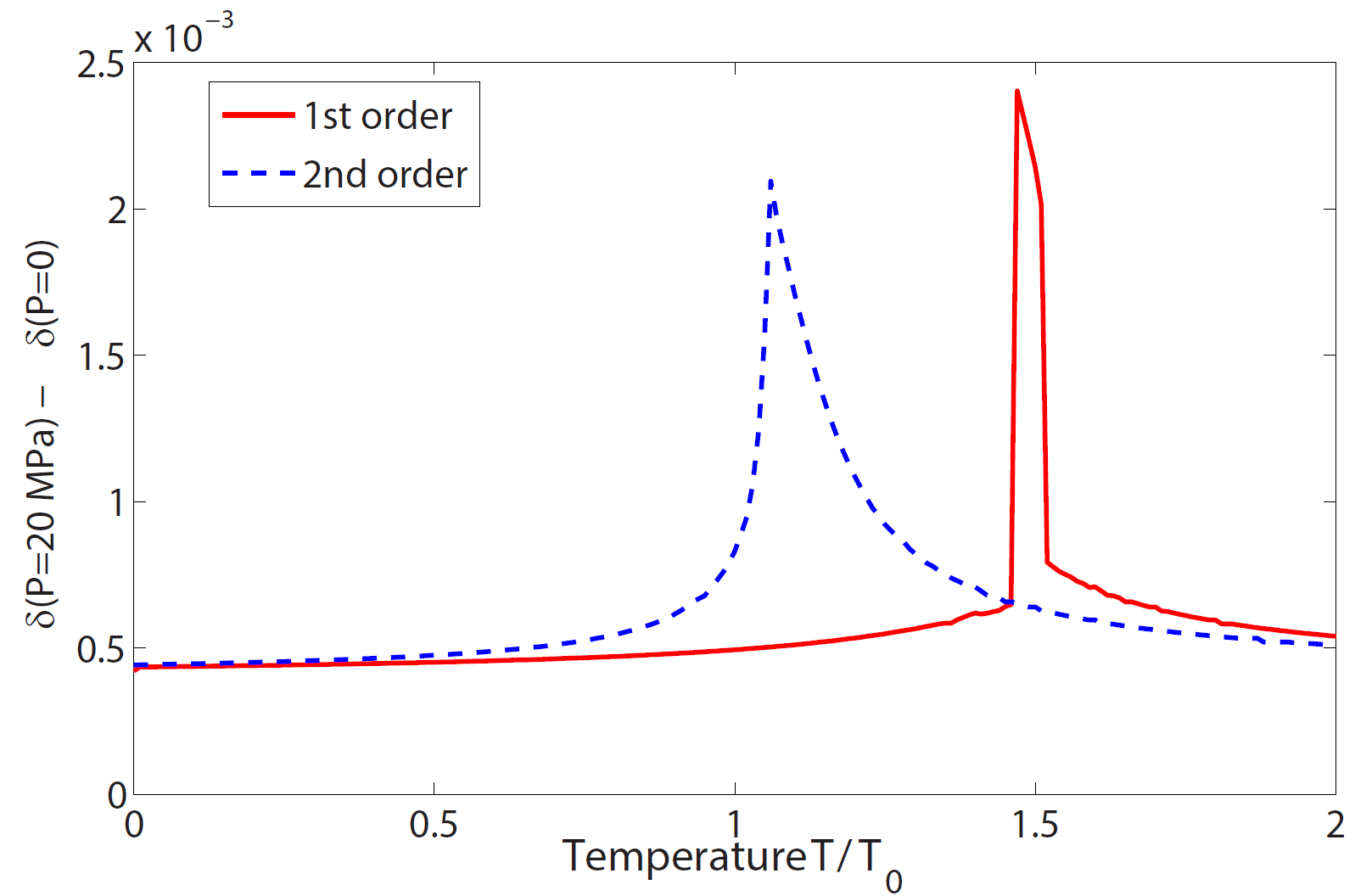}
\caption{(Color online) Change in the lattice distortion as a function of temperature, calculated from the Landau theory assuming either 1st order (Eq.~\ref{eq.first}) or 2nd order (Eq.~\ref{eq.second}) magnetic phase transition coupled to the nematic order parameter $\varphi$. The blue dashed curve is shifted to the left by $\Delta T = 0.2T_0$ for clarity. These results should be compared to the neutron data in Fig. 1c in the main text.}
\label{fig.delta-eps}
\end{figure}

Having introduced the Landau formalism above, we now study the effect of the applied external stress $P$ on the behavior of the lattice distortion. The calculated temperature dependence of $\delta(P) - \delta(0)$ is shown in figure~{\ref{fig.delta-eps}} for the realistic strain $P=20$~MPa and is shown to depend crucially on the nature of the magnetic phase transition. Indeed, the only difference between the two curves is the sign in front of the quartic $M^4$ terms in Eqs.~(\ref{eq.first}) and (\ref{eq.second}), while all the other Landau expansion parameters are kept the same (the two curves are offset horizontally for clarity). Note that for small $P$, $\delta(P) - \delta(0) \approx (\ud \delta/\ud P) P$ is proportional to the nematic susceptibility, which is expected to diverge at $T_s^{CW}$ according to Eq.~(\ref{eq.suscept}).
Both curves in Fig.~{\ref{fig.delta-eps}} exhibit an enhanced nematic susceptibility on approaching $T_s^{CW}$, as expected.
The main difference is the shape of the curve on approaching the transition, which has a distinct asymmetric ``lambda'' shape in the case of the second-order magnetic transition, and resembles closely the experimentally measured $\delta(P) - \delta(0)$ for \BaFeNi in Fig. 1c (see main text). By contrast, the N\'eel transition is first order in \SrFeNi, and the experimental behavior in Fig. 1(c) is close to the calculated sharp increase seen in our model (solid line in Fig.~{\ref{fig.delta-eps}}). Therefore, the Landau free energy results corroborate our conclusion that the nature of the magnetic transition is crucial to the observed temperature dependence of the lattice distortion.

We note in passing that for sufficiently strong  coupling constant $\mu$, the magnetic transition becomes weakly first-order even if the intrinsic free energy has positive $M^4$ term in Eq.~(\ref{eq.second}). This is likely the explanation for the observed change of the nature of the magnetic transition from weakly first order in \BaFeAs to second order upon Co doping~\cite{CLester2009, SNandi, mgkim}.
However for the values of the coupling constants in Fig.~\ref{fig.delta-eps}, this effect is imperceptible and the main difference between the two curves is due to the different intrinsic nature of the magnetic phase transition depending on the sign of the quartic term in Eqs.~(\ref{eq.first}) and (\ref{eq.second}). We have verified that for the significantly larger values of the coupling constant ($\mu\gtrsim 0.4$ in Eq.~\ref{eq.second}), it is indeed possible to obtain the shape similar to the dashed line in Fig.~{\ref{fig.delta-eps}} because the magnetic transition becomes effectively first order. In either case, our conclusions remain intact.

\subsection{Interpretation of the resistivity anisotropy}

The resistivity anisotropy $\Delta \rho = (\rho_a - \rho_b)/(\rho_a + \rho_b)$ has been widely used as a proxy for the electronic nematic order parameter in the iron pnictides~\cite{chu10,chu12}. However in some compounds, in particular in \SrFeNi, the resistivity anisotropy is vanishingly small immediately above $T_N$ [see Fig. 1(b) in the main text], whereas it is much larger in BaFe$_{2-x}T_x$As$_2$.  This is puzzling because the lattice distortion is comparable in both cases [Fig. 1(c)] and, according to Eq.~(\ref{eq.delta}), one expects the lattice distortion $\delta$ to be proportional to the nematic order parameter.

\begin{figure}[tb!]
\includegraphics[width=0.44\textwidth]{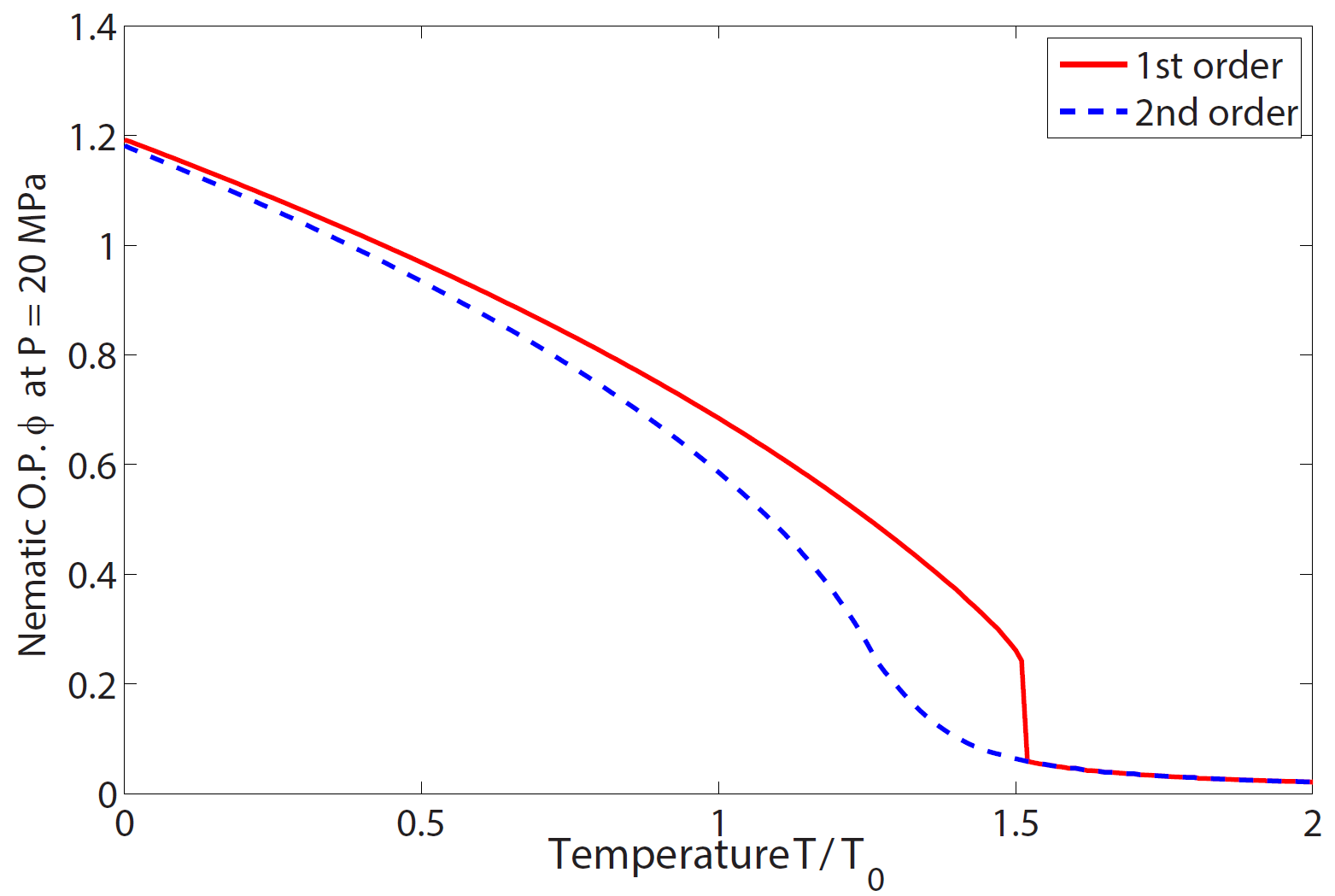}
\caption{(Color online) Calculated temperature dependence of the nematic order parameter $\varphi$ coupled to either first or second order magnetic phase transition. Note that in both cases, the bare $\varphi^4$ term is positive, however the nematic transition is rendered first order (solid curve) when coupled to the first order magnetic order parameter in Eq.~(\ref{eq.first}).}
\label{fig.phi}
\end{figure}

To shed more light on this apparent inconsistency, we have plotted in Figure \ref{fig.phi} the temperature dependence of the nematic order parameter $\varphi$ under the uniaxial stress $P=20$~MPa. The two curves correspond to the first- and second-order nature of the magnetic transition, respectively, and the Landau parameters were kept the same in both cases (except for the sign of the quartic term in Eq.~\ref{eq.first} and \ref{eq.second}). Above the transition temperature, $T>T_s$, the values of $\varphi$ are predictably small, but importantly, they are identical in the two cases. In fact, the main difference lies in the temperature dependence immediately below $T_s$. From Fig.~\ref{fig.phi}, it would appear that in this regime, the nematic order parameter should be smaller for the second-order phase transition, however this is diametrically opposite from the comparison between \BaFeNi and SrFe$_2$As$_2$ in Fig. 1(b) (see main text), where the magnetic transition in \BaFeNi is second order, yet resistivity anisotropy is much larger. This qualitative observation can be made sharper by considering Eq.~(\ref{eq.delta}), where the coupling constant $\lambda$ is estimated from experiment to be a factor of $\sim\!5$ larger in \BaFeNi and \BaFeAs compared to \SrFeNi, whereas the elastic modulus $C_{66,0}\approx 50$~GPa is similar in all three materials.
 Then, \BaFeAs is expected to have at least a factor of 5 larger lattice distortion compared to \SrFeNi, assuming that $\varphi$ is the same in both materials.
 If one now equates the resistivity anisotropy with the nematic order parameter $\varphi$, as has commonly been done in the literature~\cite{chu10,chu12}, then one is forced to conclude that $\varphi$ must be about 4 times larger in \BaFeAs due to the larger resistivity anisotropy [see Fig. 1(b)]. Taken together, one would expect the lattice distortion $\delta$ to be a factor of $\sim\!20$ larger in \BaFeAs and in \BaFeNi compared to \SrFeNi (factor of 4 due to larger resistivity anisotropy, times  factor of 5 due to larger $\lambda$).
And yet this clearly contradicts the experimental evidence in Fig.~1(c), according to which the lattice distortion is almost the same in all three materials.

One possible way out of this dilemma is that the Landau theory may not be applicable to describe the nematicity in the pnictides. However, given the excellent semi-quantitative agreement that Landau theory provides for the lattice distortion (Fig. \ref{fig.delta-eps} above) and its well documented success describing the elastic shear modulus measurements \cite{anna}, such a conclusion is perhaps not well justified. Rather, a much more plausible conclusion is that resistivity anisotropy is a poor substitute for the nematic order parameter. While it is plausible that the two quantities are proportional to each other, as follows from the nematic susceptibility measurements \cite{chu12}, the coefficient of proportionality need not be constant and can have a strong temperature dependence (and likely material dependence), as suggested recently by Tanatar \emph{et al.} in the recent study on FeSe~\cite{Tanatar15}. This material displays a non-monotonic temperature dependence of the resistivity anisotropy with a peak below $T_s$, qualitatively similar to \BaFeAs. Further theoretical and experimental studies are necessary to elucidate the precise relationship of the resistivity anisotropy and the nematic order parameter in the iron pnictides and chalcogenides. Direct microscopic measurements of the nematic order parameter, for instance using the angle-resolved photoemission spectroscopy (ARPES) to probe the orbital splitting, combined with the uniaxial pressure measurements, would be desirable.

\end{document}